\documentclass[preprint2]{aastex62}

\usepackage{amsmath}
\usepackage{hyperref}
\usepackage{listings}
\usepackage{color}
\usepackage{float}
\usepackage{xspace} 
\usepackage{longtable}

\newcommand{\stsp}{\texttt{STSP}\xspace}
\newcommand{\kepler}{\textsl{Kepler}\xspace}
\newcommand{\spitzer}{\textsl{Spitzer}\xspace}
\newcommand{\gaia}{\textsl{Gaia}\xspace}

\begin{document}

\title{Robust Transiting Exoplanet Radii in the Presence of Starspots\\ from Ingress and Egress Durations}

\author[0000-0003-2528-3409]{Brett M. Morris}

\author[0000-0002-0802-9145]{Eric Agol}
\altaffiliation{Guggenheim Fellow}

\affiliation{Astronomy Department, University of Washington, Seattle, WA 98195, USA}

\author[0000-0003-1263-8637]{Leslie Hebb}

\affiliation{Physics Department, Hobart and William Smith Colleges,  Geneva, NY 14456, USA}

\author[0000-0002-6629-4182]{Suzanne L. Hawley}

\affiliation{Astronomy Department, University of Washington, Seattle, WA 98195, USA}

\email{bmmorris@uw.edu}

\begin{abstract}
We typically measure the radii of transiting exoplanets from the transit depth, which given by the ratio of cross-sectional areas of the planet and star. However, if a star has dark starspots (or bright regions) distributed throughout the transit chord, the transit depth will be biased towards smaller (larger) values, and thus the inferred planet radius will be smaller (larger) if unaccounted for. We reparameterize the transit light curve to account for ``self-contamination'' by photospheric inhomogeneities by splitting the parameter $R_p/R_\star$ into two parameters: one for the radius ratio -- which controls the duration of ingress and egress -- and another which measures the possibly contaminated transit depth. We show that this is equivalent to the formulation for contamination by a second star (with positive or negative flux), and that it is sensitive to time-steady inhomogeneity of the stellar photosphere. We use synthetic light curves of spotted stars at high signal-to-noise to show that the radius recovered from measurement of the ingress/egress duration can recover the true radii of planets transiting spotted stars with axisymmetric spot distributions if the limb-darkening parameters are precisely known. We fit time-averaged high signal-to-noise transit light curves from {\it Kepler} and {\it Spitzer} of ten planets to measure the planet radii and search for evidence of spot distributions.  We find that this sample has a range of measured depths and ingress durations which are self-consistent, providing no strong evidence for contamination by spots. However, there is suggestive evidence for occultation of starspots on Kepler-17, and that relatively bright regions are occulted by the planets of Kepler-412 and HD 80606. Future observations with the James Webb Space Telescope may enable this technique to yield accurate planetary radii in the presence of stellar inhomogeneities.
\end{abstract}

\object{HAT-P-11, TrES-2, HAT-P-7, Kepler-17, Kepler-39, Kepler-412, GJ 1214, TRAPPIST-1}


\section{Introduction}

Precise and accurate radii of exoplanets are critical to understanding their habitability. Radii are often used as a proxy for bulk composition when masses aren't available, informed by population studies which suggest that planets smaller than $R\lesssim1.5R_\oplus$ are more likely rocky than gaseous \citep{Weiss2014,Rogers2015,Fulton2017}. Since bulk densities are sensitive to the radius to the third power, accurate radii are required in order to infer accurate bulk densities. 

The radii of transiting exoplanets are typically constrained by the transit depth, which is interpreted as the ratio of the cross-sectional areas of the planet and star $\Delta \mathcal{F}/\mathcal{F} = (R_p/R_\star)^2$ (ignoring limb-darkening). The transit depth accuracy as a measure of the planet's radius, however, is sensitive to stellar surface inhomogeneities. When planets pass over dark star spots, for example, the transit light curve is biased towards shallower transit depths during the transit \citep[see e.g.][]{Csizmadia2013,Oshagh2013}. As we discover more planetary systems of late type stars like TRAPPIST-1 \citep{Gillon2016,Gillon2017,Luger2017}, the time- and wavelength-dependent variability of these active stars and non-uniformity is likely to limit the accuracy and precision of our measurements of the planet radii, especially as a function of wavelength \citep{Roettenbacher2017,Rackham2018,Zhang2018,Morris2018c}.

Particularly insidious are nearly axisymmetric stellar inhomogeneities and/or 
inhomogeneities which are comprised of odd harmonics.  These do not lead
to variability in the total flux of the star, and do not necessarily
cause spot-crossing features in the transit light curve;  yet they will 
still  affect the depth of transit, and hence the inferred planet-star 
radius ratio and transmission spectrum.
Consequently, even without indications of star-spot activity in the light
curve of a star, spots (or other inhomogeneities) can still affect
the depths of transits;  this phenomenon is referred to as the ``transit
light source effect" in the recent work by \citet{Rackham2018}. These forms
of time-steady ``activity" might be
due to a multitude of star spots (such as along a particularly active 
latitude), due to gravity darkening, or due to a strong global magnetic field.
The message of this paper is that if one measures the duration of ingress 
as compared to the total transit
duration, this time-dependent quantity will be less affected by the presence
of spots on the rest of the photosphere, and thus may give a more
precise, albeit less accurate, measurement of the planet-star radius ratio
compared with measuring the radius ratio from the depth of transit.
However, in most cases it does require an independent constraint upon the 
impact parameter of the transit, as well as a careful handling of limb-darkening.

In Section~\ref{sec:rederiving} we reparameterize the transit light curve formalism of \citet{Mandel2002} for more robust planet radii in the presence of significant starspot coverage, and elaborate on the degeneracies in the model in Section~\ref{sec:theory}. In Section~\ref{sec:valid}, we validate the new transit model with fits to synthetic light curves of simulated spotted stars. We fit real transit light curves of a variety of planets including TRAPPIST-1 and GJ 1214 in Section~\ref{sec:app}, and discuss the results and conclusions in Sections~\ref{sec:discussion} and \ref{sec:conclusion}.

\section{Self-contamination and the transit light curve} \label{sec:rederiving}

Following \citet{Mandel2002}, the ratio of unocculted flux to occulted flux $F$ is given by 
\begin{equation}
F^e(p_0,z,u_1,u_2) = 1-\lambda^e(p_0,z,u_1,u_2),
\end{equation}
where $p_0 = R_p/R_\star$, $z$ is the projected sky separation between the centers of the star and planet (in units of the stellar radius), and $u_1$ and $u_2$ are the quadratic limb-darkening parameters. One way to interpret the transit light curve -- for a star known to have little contamination from other nearby stars -- is to create a flux contamination term which will account for bright or dark regions of the surface of the star known to be transited. We introduce the new parameter $p_1$ and renormalize the $\lambda$ term as:
\begin{equation} \label{eqn:robin}
F^e(p_1,p_0,z,u_1,u_2)= 1-\left(\frac{p_1}{p_0}\right)^2 \lambda^e(p_0,z,u_1,u_2).
\end{equation}
It follows that $p_1 \approx \sqrt{\delta}$ where $\delta$ is the transit depth (the approximation becomes an equivalence for the uniform case where $u_1=u_2=0$). We emphasize here that $p_1$ is not a radius ratio parameter like $p_0$; $p_1^2$ is related to the observed transit depth in the presence of bright or dark spot contamination. A measured difference between $p_0$ and $p_1$ is in principle sensitive to starspots, even with an axisymmetric distribution, unlike rotational modulation.

This renormalization is equivalent to the analysis often performed on \kepler planet candidate host stars in order to search for contaminating light from a second, unresolved star \citep{Torres2011,Morton2012,Teske2018}. Typically in the external contamination case, the light from the background star dilutes the flux of the exoplanet host star, decreasing the transit depth, modifying the inferred impact parameter, and leading to underestimated the planet radii, if the contamination is unaccounted for.

In equation \ref{eqn:robin}, we are essentially allowing for a contaminating light source on the host star itself, which can be positive \textit{or} negative due to bright or dark regions on the stellar surface, thus increasing or decreasing the transit depth $p_1^2$ with respect to the expectation given the planet's radius $p_0$. Typically for contaminating nearby sources, the ratio of the unocculted flux to occulted flux is given by 
\begin{equation}
F^e = (1 - \lambda^e)(1 - \epsilon) + \epsilon,
\end{equation}
where $\epsilon$ is the flux of the contaminating source. In this parameterization, 
\begin{equation}
\left( \frac{p_1}{p_0}\right)^2 = 1 - \epsilon
\end{equation}
We focus our attention in this work on the ``self-contamination'' effect of bright or dark regions on the host star itself. 

We have implemented the algorithm in CPython called \texttt{robin}, based on a fork of the \texttt{batman} code by \citet{Kreidberg2015}, which is publicly available\footnote{Open source, available online: \url{https://github.com/bmorris3/robin}}.

\section{Constraining the radius ratio from durations of the transit and of ingress/egress} \label{sec:theory}

Given the untoward effect of contamination on measurement of the planet-star radius ratio,
we require some additional constraint to derive this ratio in the presence of a
heterogenous stellar photosphere or blending with other sources of light.  This
can be provided by a purely geometrical constraint:  the ratio of ingress duration
to transit duration.  However, to implement this requires a knowledge of (or constraint
upon) the impact parameter of the transit, $b$, which we describe in this section.

The duration of ingress for an eccentric orbit  
is given approximately by:
\begin{eqnarray}
 \tau = \frac{2R_p}{v\sin{\theta}},
\end{eqnarray}
where $\theta$ is the angle between the tangent of the limb of the star and the path of
the planet (projected onto the sky plane) and $v$ is the sky velocity of the planet
relative to the star during transit, which we take to be constant in this section.
The duration of the transit, $T$,
from mid-ingress to mid-egress is given by
\begin{equation}
 T = \frac{2R_*\sin{\theta}}{v},   
\end{equation}
Now $\sin{\theta} = \sqrt{1-b^2}$, so
\begin{eqnarray}
 \tau &=& 2(1-b^2)^{-1/2}\frac{R_p}{R_*}\frac{R_*}{v},\\
  T &=& 2(1-b^2)^{1/2}\frac{R_*}{v}.  
\end{eqnarray}
These formulae assume that $a \gg R_*$ and $R_p \ll R_*$, neglecting the curvature
of the orbit and the stellar limb (see \citet{Winn2010} or the appendix for more 
complete relations).

The transit and ingress/egress durations are simply a function of time, and
do not depend upon the transit depth, and hence will not be affected by dilution
by star spots or flux from a blend, such as a companion star.  Thus, a precise
measurement of $\tau$ and $T$ can in principle give another means of constraining
the radius ratio of the planet to the star which will be less affected by
blending or a heterogeneous stellar photosphere.

There are two ways to solve for this radius ratio, the first of which is
derived from the ratio of ingress to duration of transit,
\begin{eqnarray}
 \frac{R_p}{R_*} = \frac{\tau}{T}(1-b^2). \label{eqn:rprs}
\end{eqnarray}
This requires a constraint upon the impact parameter, $b$.  The {\it shape}
of ingress and egress depends weakly on the impact parameter;  however, this
is generally too subtle to measure, even with the highest signal-to-noise
transits.  Note that a given measured value of $\tau$ and $T$ imply a
maximum value of $R_p/R_*$, which is the value this ratio would have
if $b=0$.  This could be useful for placing upper limits on the radii of
planets given measured properties of the transit;  this constraint is
independent of the eccentricity and period of an orbit and independent of the
density of the star.  If the depth of transit (as parameterized by
$p_1^2$) implies a radius ratio that is greater than this maximum value,
then this likely implies that both a small impact parameter and that non-transited
star spots must be present (or that the planet transits a bright chord)
to cause a larger transit depth, or that the planet is significantly oblate 
(see next section).

The second means of solving for the radius ratio from time-dependent
quantities is given by
\begin{equation}
\frac{R_p}{R_*} = \frac{\tau T}{4} \left(\frac{v}{R_*}\right)^2,
\end{equation}
which requires a measurement of the normalized sky velocity, given by:
\begin{eqnarray}
\frac{v}{R_*} = \frac{2\pi}{P}\frac{a}{R_*}\frac{1+e\sin{\omega}}{\sqrt{1-e^2}},\\
\frac{a}{R_*} = \left[\frac{G\rho_*P^2}{3\pi}\right]^{1/3},
\end{eqnarray}
where $\rho_*$ is the stellar density, $e$ is the orbital eccentricity, and
$\omega$ is the longitude of periastron of the star measured along the orbital path from
when the star crosses the sky plane away from the observer.\footnote{For nearly
edge-on orbits, transits
occur when the true longitude of the star is $\theta=\pi/2$.  The relative
velocity of the star and planet is maximum at periastron, and if the time of
periastron coincides with the time of transit, then $\omega = \pi/2$, at
which time the velocity is $\sqrt{(1+e)/(1-e)}$ of the mean orbital velocity.}
Thus, if $a/R_*$ is constrained, for example, from asteroseismology or asterodensity
profiling (which requires additional transiting planets), and if the
eccentricity vector, $\vec{e}=\{e\cos{\omega},e\sin{\omega}\}$ is constrained, 
for example, from radial velocity measurements
or transit-timing variations, then
the transit chord, $d/R_*=2\sqrt{1-b^2}$ (in units of $R_*$), and hence the impact
parameter, can be constrained from the transit duration.

In practice, then, the radius ratio may only be determined in cases for which
there is a prior on the impact parameter, and generally this is derived from
the dependence of the transit duration upon the stellar density and the eccentricity
of the planet's orbit, as well as other well-measured quantities such as the
planet's orbital period.  The formulae above are only approximate as they neglect
the curvature of the orbit and the curvature of the limb of the star (see the
Appendix~\ref{sec:simpleeqns} for more general expressions), but these
relations are generally good approximations, while the full relation can be
accounted for with a full orbital transit model.

\subsection{Planetary Oblateness}

The preceding analysis assumes that a planet is spherical.  However, the
transit depth and the duration of ingress and egress are also affected by
the shape of a planet.  If a planet is distorted, for example, due to 
oblateness, then the area of a planet and the distance a planet moves 
during ingress and egress are not simply related to the mean radius of the planet.
Oblateness can be induced by rotation of a planet, causing a bulging of
the equator due to centripetal acceleration \citet{Barnes2003}, or by thermal structure 
of a planet, causing a larger scale height at the hotter equator
relative to the poles \citep{DobbsDixon2012}.  
There are several references in the literature which fully consider the 
transits of oblate planets, and we refer the interested reader to  
\citet{Hui2002,Barnes2003,Carter2010,Zhu2014,Biersteker2017}. 

In addition to bright or dark active regions and contaminating light sources, planetary oblateness can cause a mismatch between the observed values of $p_0$ and $p_1$. Here we derive a relationship between $p_0, p_1$, and the oblateness of the planet,
\begin{equation}
    f = \frac{R_\mathrm{eq} - R_\mathrm{pole}}{R_\mathrm{eq}}.
\end{equation}

\begin{figure}
    \centering
    \includegraphics[scale=0.9]{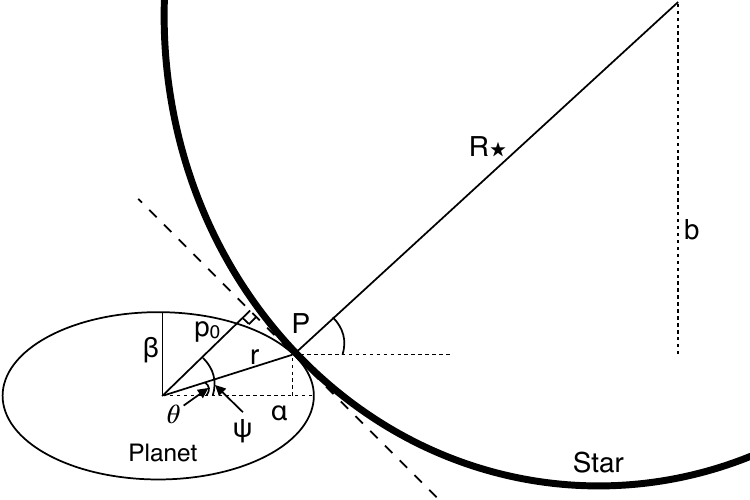}
    \caption{Oblate planet model, for a planet with projected semimajor axis $\alpha$ and semiminor axis $\beta$.}
    \label{fig:oblate}
\end{figure}

The projected ellipsoid of an oblate planet with semimajor axis $\alpha$ (in units of $R_\star$) and semiminor axis $\beta$ makes first contact with the star at point $P$, see Figure~\ref{fig:oblate}, which is an angle $\theta$ from the semimajor axis of the planet (or the $+\hat{X}$ axis in the planet-centered coordinate system). The angle from the $+\hat{X}$-axis to the normal of the tangent of the ellipse $\phi = \sin^{-1}(b)$, where $b$ is the impact parameter. 

The definition of the elliptical outline of the planet $r$ is
\begin{equation}
r^2 = \alpha^2 \cos^2 \theta + \beta^2 \sin^2 \theta,
\end{equation}
and the slope of a tangent line which meets the ellipse at point $P$ is
\begin{equation}
m = -\frac{\beta}{\alpha} \cot{\theta} = -\cot{\psi}.
\end{equation}
We assume for simplicity the major axis of the oblate planet is
aligned along its direction of motion relative to the star, so $\alpha = R_{eq}/R_*$
and $\beta = R_{pole}/R_*$
Now we can solve for $p_0$ given 
\begin{equation}
p_0 = r \cos{(\psi-\theta)}
\end{equation}
where $\psi$ is the angle measured up from the $+\hat{X}$ axis to $p_0$. Solving for $p_0$ in terms of $\alpha, \beta,$
\begin{equation}
p_0 = \alpha \frac{\sqrt{1-b^2+b^2(1-f)^4}}{1-b^2+b^2(1-f)^2} (1-b^2f)
\end{equation}
Note that in the oblate planet case $f > 0$, $p_0 \neq \alpha$ for non-zero $b$. The depth of the transit will be the ratio of the cross-sectional area of the planet to the star, 
\begin{equation}
p_1^2 = \alpha \beta,
\end{equation}
so the general form of the relation between $p_0,p_1,b$ and $f$ is:
\begin{eqnarray}
\left(\frac{p_1}{p_0}\right)^2 &=&  \frac{(1-f) (1-b^2+b^2(1-f)^2)^2}{(1-b^2f)^2(1-b^2+b^2(1-f)^4)}\\
&\approx& 1-f (1-2b^2),
\end{eqnarray}
where the last expression is valid for $f \ll 1$.
In the special case where $b=0$, this simplifies to 
\begin{equation}
\left(\frac{p_1}{p_0}\right)^2 = 1-f,
\end{equation}
implying that $p_0^2 = \alpha^2 = R_\mathrm{eq}^2/R_*^2$ and $p_1^2 = \alpha\beta = R_\mathrm{pole}R_{eq}/R_*^2$.

\section{Validation on simulated light curves} \label{sec:valid}

To estimate the ability to measure the planet-star radius ratio in the presence of a heterogeneous stellar photosphere, we used the reparameterized transit light curve to fit simulated light curves generated with \stsp (Hebb et al.~2018\footnote{Open source, available online: \url{https://github.com/lesliehebb/stsp}}). \stsp synthesizes transit light curves of spotted stars by analytically computing the overlap between the planet, starspots, and the star, with limb darkening approximated by concentric circles of constant surface brightness. 

We consider two axisymmetric starspot distributions: the first in which the planet's orbital angular momentum is aligned with that of the star in the sky plane, and the planet crosses an active latitude with dark spots, and the second in which active latitudes are present with dark spots, but not transited by the planet.  This first case causes shallower transits relative to a uniform star, which is equivalent to contamination by an additional source with $\epsilon > 0$, while the second case causes deeper transits, and is equivalent to dilution by a source with {\it negative} flux, $\epsilon < 0$, which is the deficit of stellar flux caused by the darker spots.

\subsection{Occultation of an active latitude} \label{sec:activelat}

\begin{figure*}
    \centering
    \includegraphics[scale=0.5]{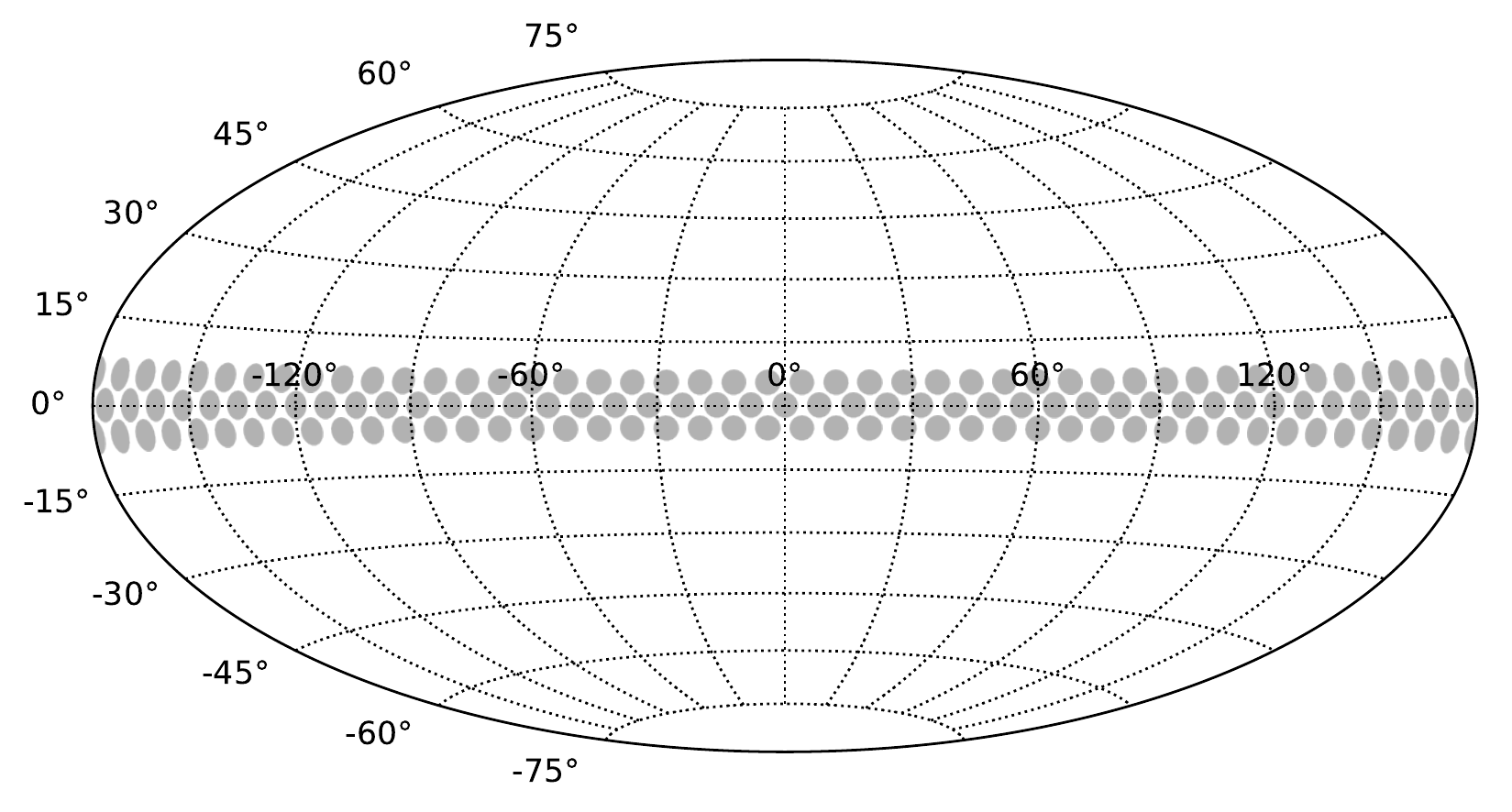}
    \includegraphics[scale=0.6]{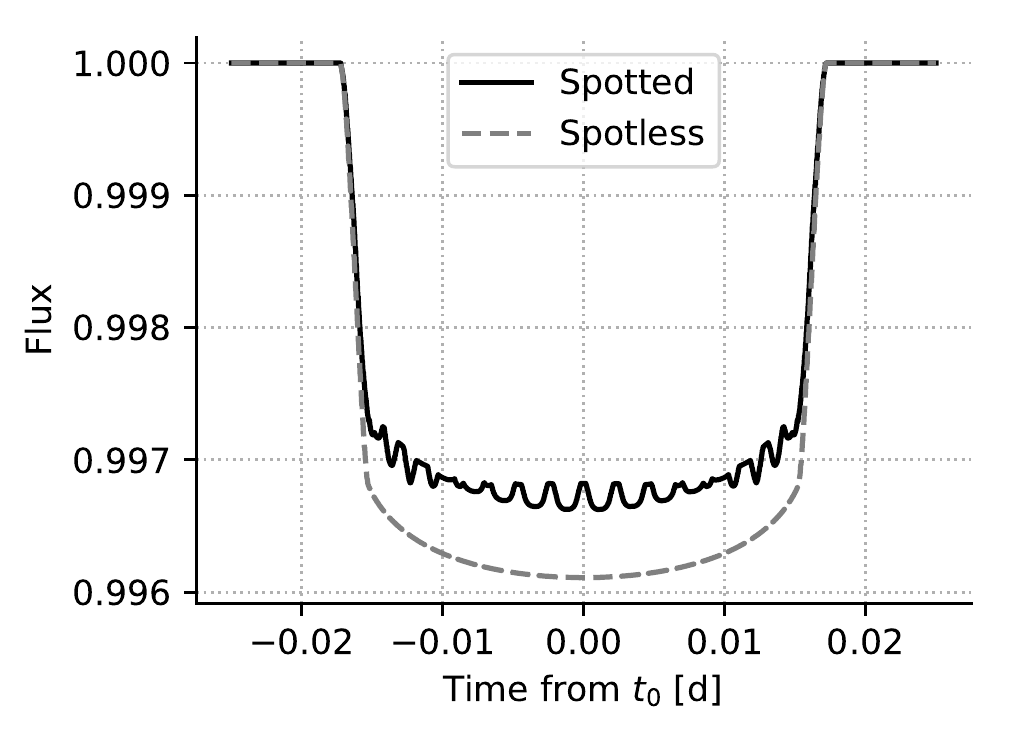}
    \caption{\textsl{Left:} Hypothetical spot map for a dense band of spots in an active latitude at the stellar equator. \textsl{Right:} Transit of a small planet across the star with the spot map on the left, with $i_o = 90^\circ$ and $\lambda = 0^\circ$ and other parameters set to those of TRAPPIST-1 g (black curve), compared with the transit of the same system without spots (gray dashed curve). At lower S/N, or for different spot geometries, the bottom of the spotted transit might simply appear flat and relatively shallow, but the ingress and egress durations are the same for both light curves, allowing us to recover the true planet radius from timing, independent of the transit depth which is affected by starspots.}
    \label{fig:map}
\end{figure*}

\begin{figure*}
    \centering
    \includegraphics[scale=0.75]{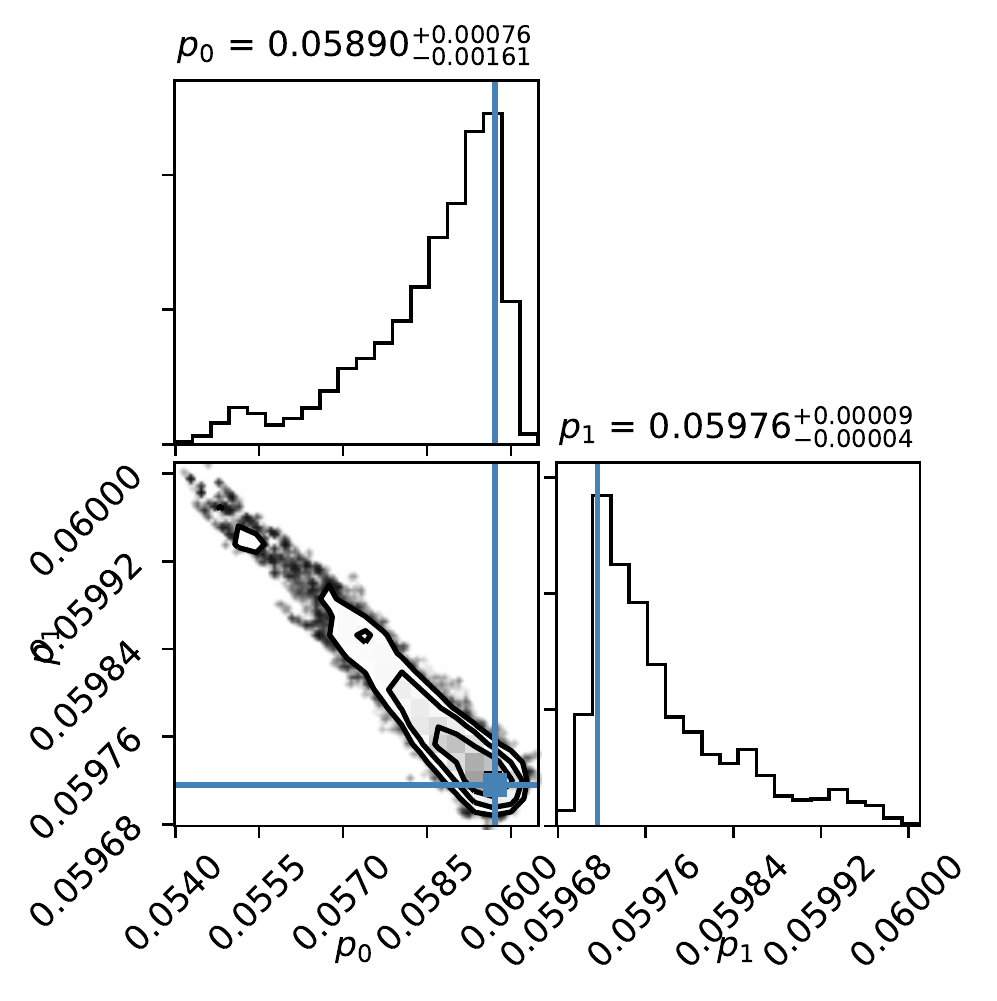}
    \includegraphics[scale=0.75]{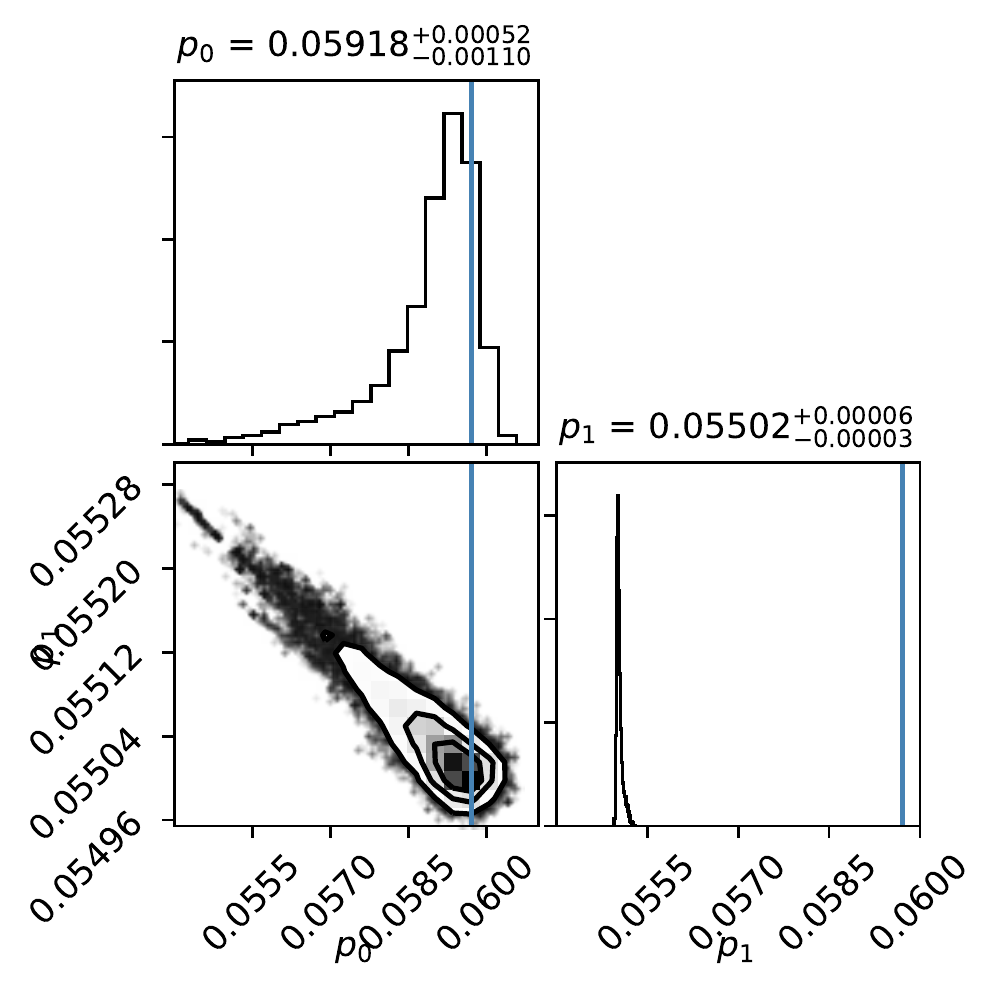}
    \caption{\textsl{Left:} posterior distributions from a fit to the spotless transit light curve in Figure~\ref{fig:map}, showing agreement between the radius measurement from the ingress/egress duration, $p_0 = R_p/R_\star$, and the radius measurement from the transit depth $p_1 \approx \sqrt{\delta}$ where $\delta$ is the transit depth. \textsl{Right:} posterior distributions from a fit to the spotted transit light curve in Figure~\ref{fig:map}, demonstrating $p_0 > p_1$, and that the duration-dependent radius measurement $p_0$ recovers the true radius (vertical blue line).}
    \label{fig:corners}
\end{figure*}

We synthesize a transit light curve of a star with a dense band of spots that blankets the stellar equator using \stsp with $c=0.7$ -- see Figure~\ref{fig:map}. We observe the transit at one second cadence with uncertainties of 48 ppm for each flux measurement, which is similar in scale to the oscillations in flux at the ``bottom'' of the transit light curve, such that fits to the transit light curve with the standard \citep{Mandel2002} model yield a reduced-$\chi^2 = 0.95$. This is a very high signal-to-noise example, even exceeding reasonable S/N estimates for bright targets observed with JWST. The corresponding transit light curve for a planet with impact parameter $b=0$ and projected spin-orbit alignment $\lambda = 0^\circ$ (i.e.~the planet transits the stellar equator) is \textit{shallower} than the transit of the same system without starspots. The duration of ingress and egress independently encode the radius of the planet. 

Figure~\ref{fig:corners} shows the posterior distributions of fits for the parameters $p_0, p_1$ for the spotless (left) and spotted (right) light curves, allowing the impact parameter and transit duration $T_{14}$ to vary, and holding the limb darkening parameters fixed at their true values. In the spotless light curve case, as expected, we find that $p_0 = p_1$. In the case of transiting an active latitude of dark spots, $p_0 > p_1$, indicating that the transit depth is  being decreased by occultations of dark starspots. The solution for $p_0$ is consistent with the injected transit radius $p_0=R_p/R_\star = 0.05971$. 

The degeneracy between $p_0$, the limb-darkening parameters, and the impact parameter becomes increasingly important as one decreases the S/N of the observations. For example, we inflated the uncertainties on each observation by a factor of three, and we found that the uncertainties on the $p_0$, impact parameter, and limb-darkening parameters increased by factors of 5, 2, and 2, respectively. This exercise demonstrates the importance of having independent measurements of the impact parameter or $a/R_\star$, as discussed in Section~\ref{sec:theory}.

\subsection{Occultation of a bright latitude} \label{sec:brightlat}

\begin{figure*}
    \centering
    \includegraphics[scale=0.5]{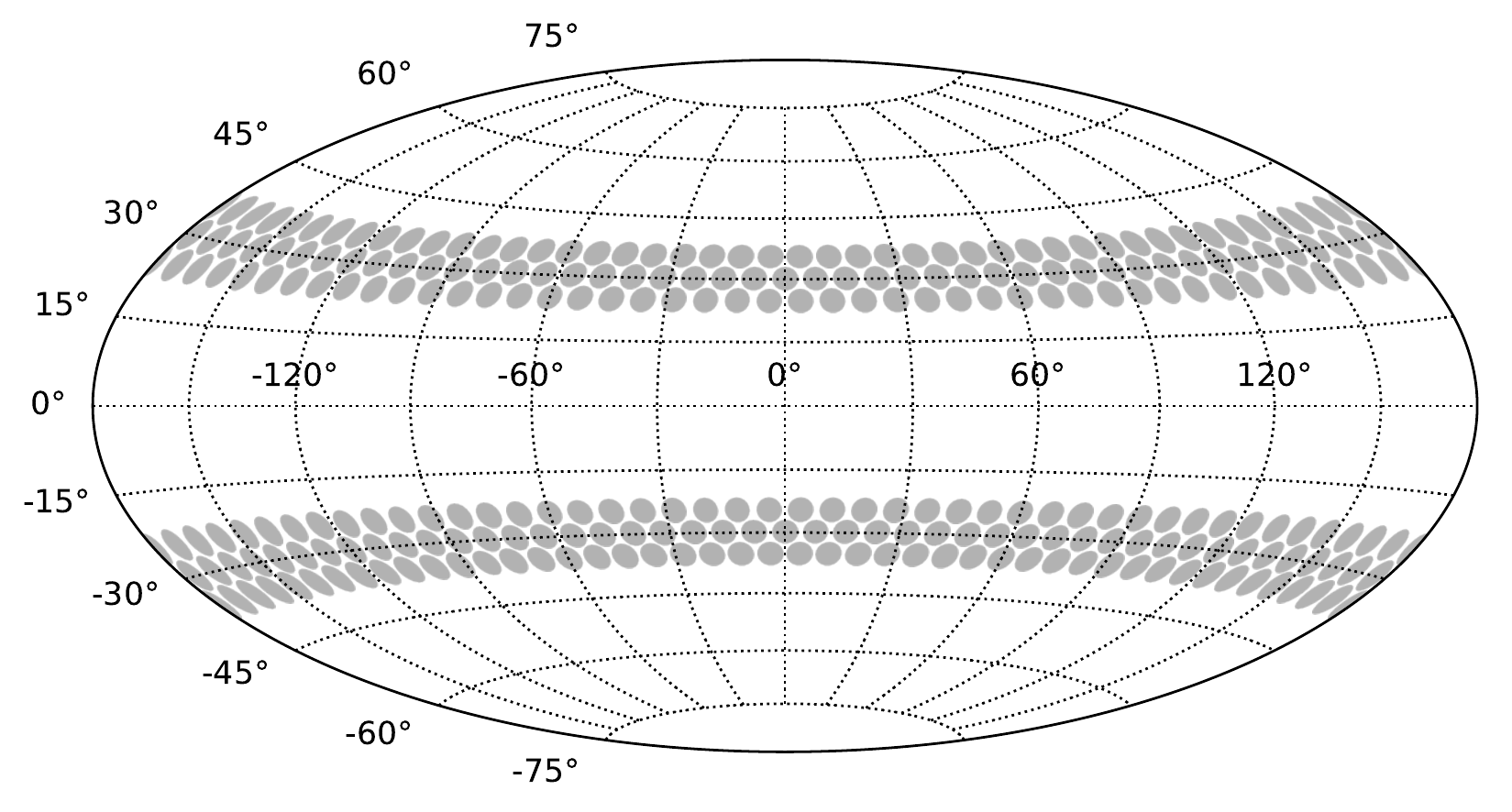}
    \includegraphics[scale=0.6]{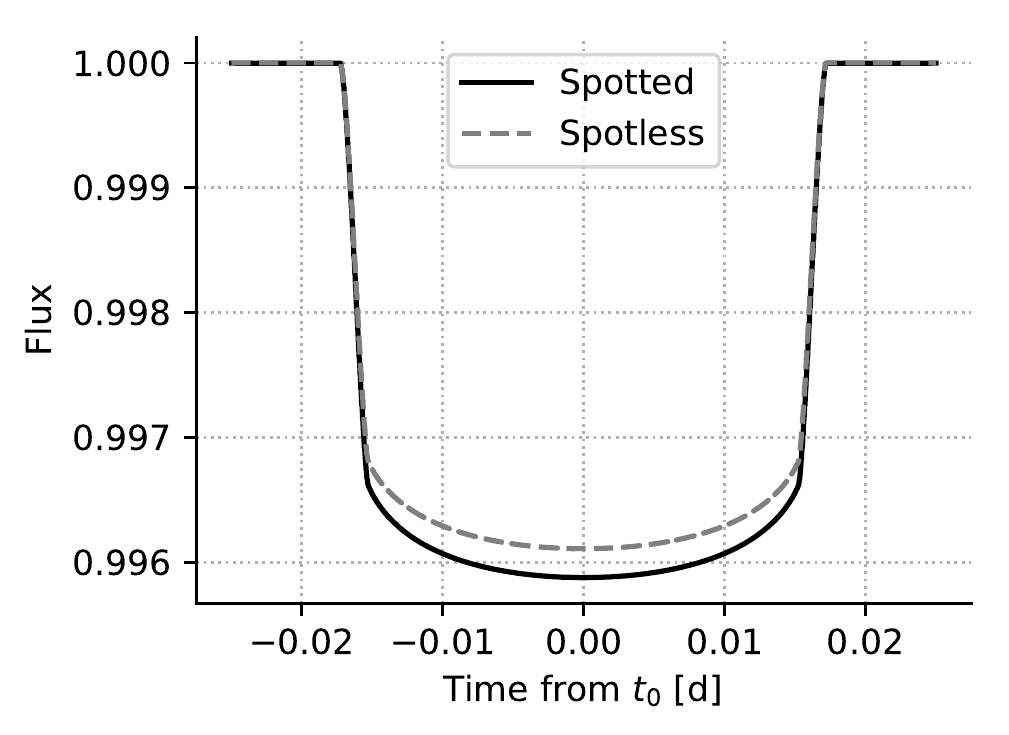}
    \caption{\textsl{Left:} Hypothetical spot map for two dense active latitudes of spots centered on $\pm 30^\circ$ latitude. \textsl{Right:} Transit of a small planet across the star with the spot map on the left, with $i_o = 90^\circ$ and $\lambda = 0^\circ$ and other parameters set to those of TRAPPIST-1 g (black curve, \citealt{Delrez2018}), compared with the transit of the same system without spots (gray dashed curve). The bottom of the spotted transit might simply appear deeper, but the ingress and egress durations are the same for both light curves, allowing us to recover the true planet radius from timing, independent of the transit depth which is affected by starspots.}
    \label{fig:map2}
\end{figure*}

\begin{figure*}
    \centering
    \includegraphics[scale=0.75]{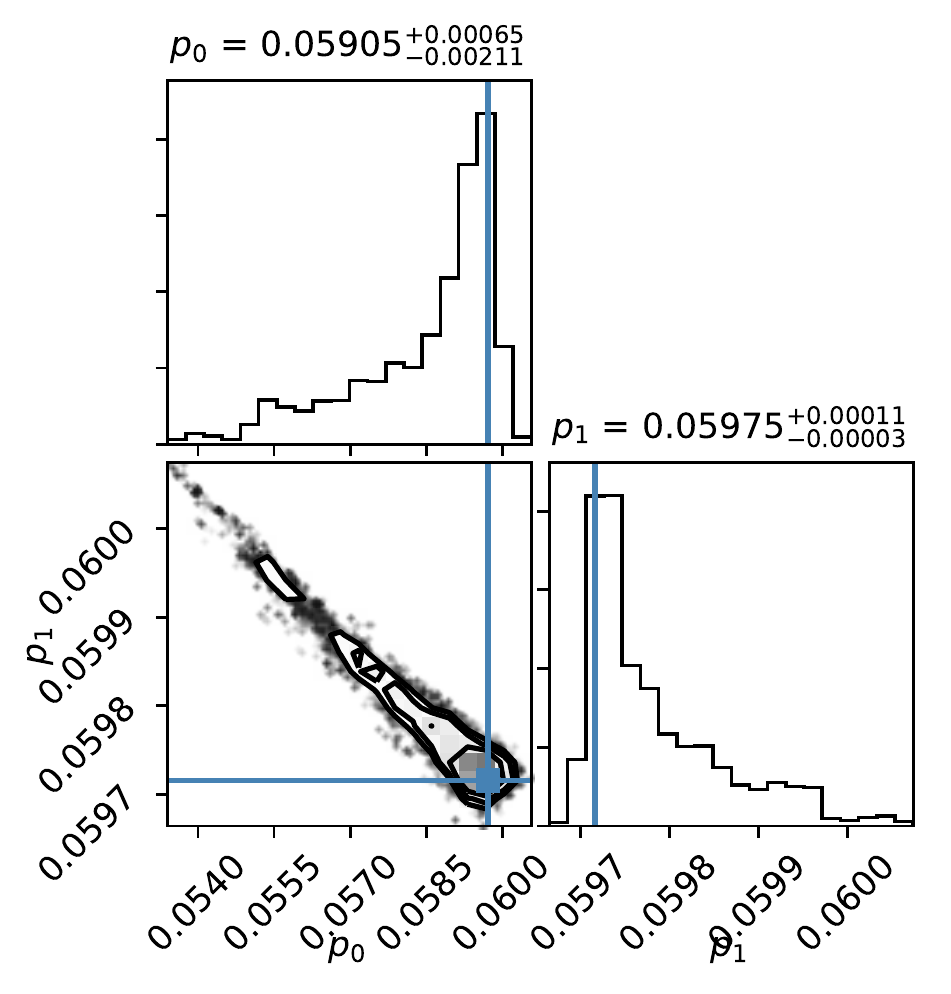}
    \includegraphics[scale=0.75]{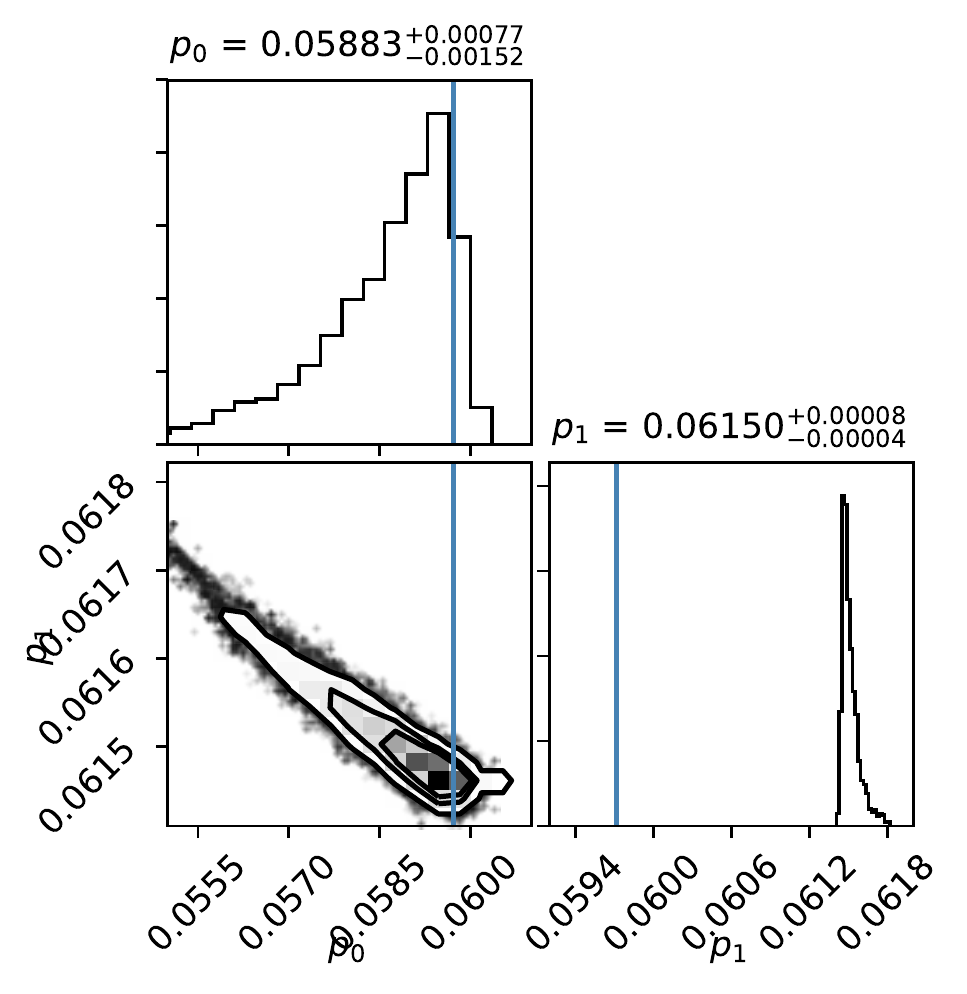}
    \caption{\textsl{Left:} posterior distributions from a fit to the spotless transit light curve in Figure~\ref{fig:map2}, showing agreement between the radius measurement from the ingress/egress duration, $p_0 = R_p/R_\star$, and the radius measurement from the transit depth $p_1 \approx \sqrt{\delta}$ where $\delta$ is the transit depth. \textsl{Right:} posterior distributions from a fit to the spotted transit light curve in Figure~\ref{fig:map2}, demonstrating $p_0 < p_1$, and that the duration-dependent radius measurement $p_0$ recovers the true radius (vertical blue line).}
    \label{fig:corners2}
\end{figure*}

We approximate a bright latitude centered on the stellar equator with \stsp by creating active latitudes of spots centered on $\pm30^\circ$ latitude, see Figure~\ref{fig:map2}, while keeping the same planet properties and uncertainties as in Section~\ref{sec:activelat}. The planet transits a relatively bright region of the stellar surface, so in this example the transit light curve for the spotted star is \textit{deeper} than expected for the spotless case. 

Figure~\ref{fig:corners2} shows the posterior distribution of fits for the parameters $p_0, p_1$ for the spotless (left) and spotted (right) light curves, allowing the impact parameter and transit duration $T_{14}$ to vary, and holding the limb darkening parameters fixed at their true values, as in Section~\ref{sec:activelat}. In the spotless light curve case, we again find that $p_0 = p_1$. In the case of transiting a bright latitude, $p_0 < p_1$, indicating that the transit depth is being increased by the spot distribution. The solution for $p_0$ is again consistent with the injected transit radius $p_0=R_p/R_\star = 0.05971$. 

We consider further limiting cases with extreme spot distributions in Appendix~\ref{sec:limitingcases}, which suggests that plausible limits for self-contamination are $-0.5 \lesssim \epsilon \equiv 1 - (p_1/p_0)^2 \lesssim 0.5$.

\subsection{Required Precision}

It is important to note that extremely high S/N light curves are required in order to produce meaningful constraints on the difference between $p_0$ and $p_1$, similar to the S/N assumed in the previous two subsections. Descriptions of the uncertainties on $p_0$ and $p_1$ in terms of planetary system and observing parameters are detailed in the Appendix. Here we will take the examples of the previous two subsections and ask what photometric precision is necessary to measure significant deviations between $p_0$ and $p_1$. 

\begin{figure*}
    \centering
    \includegraphics[scale=0.9]{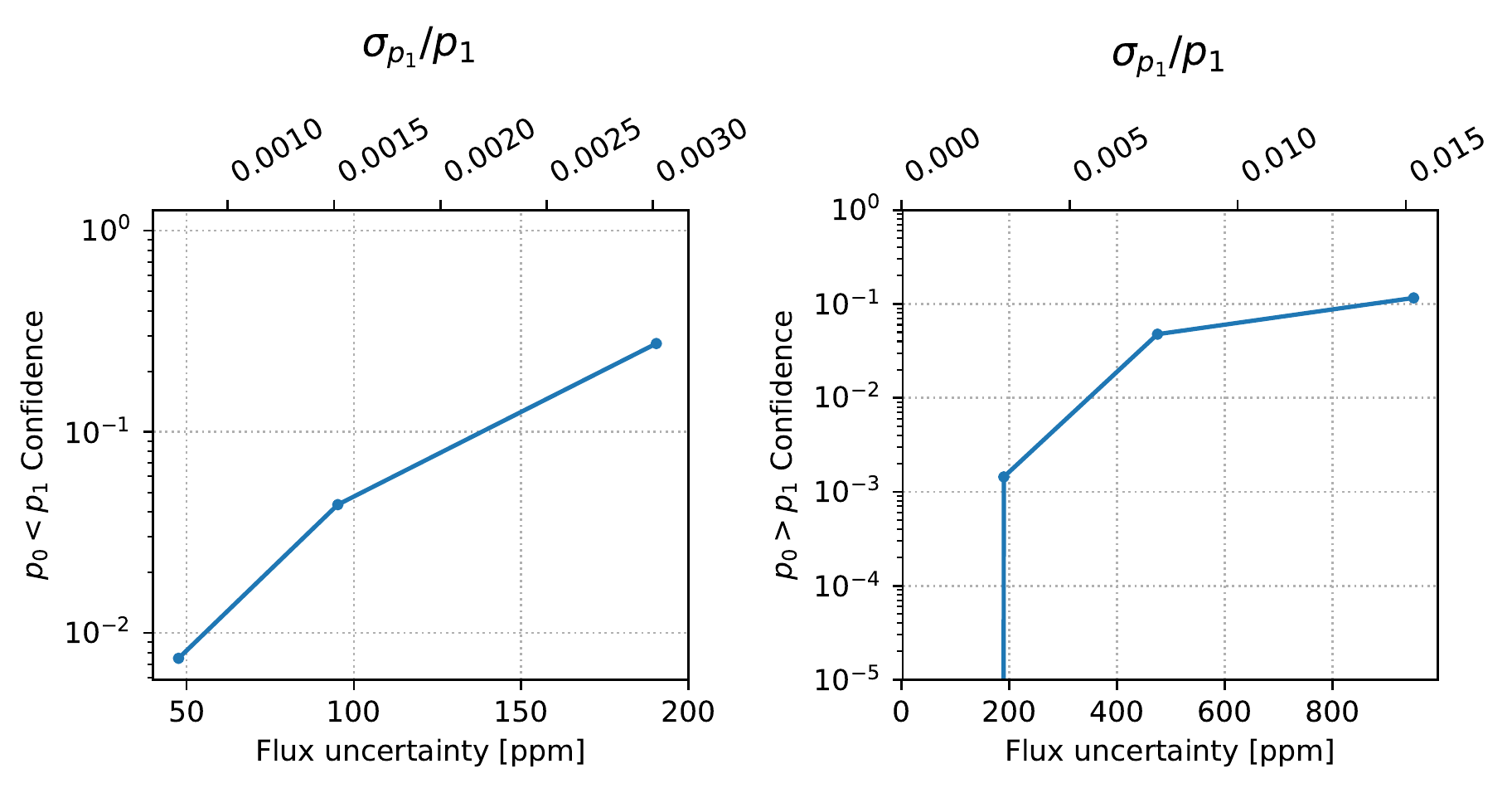}
    \caption{Posterior probabilities -- for $p_0 < p_1$ given the spot distribution in Section~\ref{sec:activelat} (crossing a dark latitude, {\it left}), and for $p_0 > p_1$ given the spot distribution in Section~\ref{sec:brightlat} (crossing a bright latitude, {\it right}) -- as a function of the photometric uncertainty on one-second cadence observations of TRAPPIST-1 g. Photometric uncertainties must be smaller than 50 ppm for a significant detection (3$\sigma$) of the spot distribution in Section~\ref{sec:activelat}, or smaller than 300 ppm for a significant detection of the spot distribution in Section~\ref{sec:brightlat}. The upper axis translates the flux uncertainty into relative uncertainty on the $p_1$ parameter.
    }
    \label{fig:uncertainty}
\end{figure*}

Figure~\ref{fig:uncertainty} shows the confidence interval for detecting $p_0 < p_1$ when crossing a dark latitude with the spot map from Section~\ref{sec:activelat}, as a function of photometric precision on one-second cadence observations of TRAPPIST-1 g (left); and the confidence interval for detecting $p_0 > p_1$ when crossing a bright latitude with the spot map from Section~\ref{sec:brightlat} (right). For the TRAPPIST-1 g system in particular, extremely high S/N precision -- 10s of ppm uncertainty per flux measurement --  would be required to significantly detect the occultation of a dark latitude. The situation is a bit less severe for the occultation of a bright latitude -- 100s of ppm uncertainty per flux measurement are sufficient to significantly detect $p_0 < p_1$ in this case. In each case, we note that the required precision is similar to or better than the precision achievable with phase-folded \kepler observations, so we anticipate that any detections of spot distributions with the self-contamination technique from \kepler observations (Section~\ref{sec:kepler}) will yield only marginal significance, and higher S/N observations, like those of JWST, may be able to achieve the precision necessary to unambiguously separate the posterior distributions of $p_0$ and $p_1$. 

\section{Application to Observations} \label{sec:app}
\subsection{\kepler Light Curves} \label{sec:kepler}

We apply the new parameterization to the highest S/N transit light curves available -- those of the transiting planets Kepler-1, Kepler-2 and Kepler-3 (a.k.a. TrES-2, HAT-P-7 and HAT-P-11). Each of these light curves have short cadence observations throughout the \kepler mission, short period planets, and very bright host stars ($K_p=9.84, 9.33, 7.00$, respectively). As a result, these light curves should yield some of the best constraints on ingress and egress durations available. We also explore some exceptional cases: the highly spotted star Kepler-17, the potentially oblate planet Kepler-39 b, and a planet with a vary high impact parameter, and thus long ingress/egress durations, Kepler-412.

For Kepler-1 through -3, the Kepler Input Catalog estimated contamination in the \kepler aperture is $\lesssim 1\%$, so a third light source can likely be ruled out as the cause for discrepancies between $p_0$ and $p_1$ for these targets \citep{Brown2011}. For Kepler-17, -39 and -412, the contamination is $\lesssim 6\%$, so discrepancies larger than $6\%$ are required to invoke stellar activity as the culprit for disagreement between $p_0$ and $p_1$.

For each \kepler transit light curve, we normalize the SAP flux by a quadratic fit to the out-of-transit flux. We assume the periods, midtransit epochs, eccentricities and arguments of periastron for each target listed in the NASA Exoplanet Archive. 


\subsubsection{TrES-2 (Kepler-1)}

\begin{figure*}
    \centering
    \includegraphics[scale=0.75]{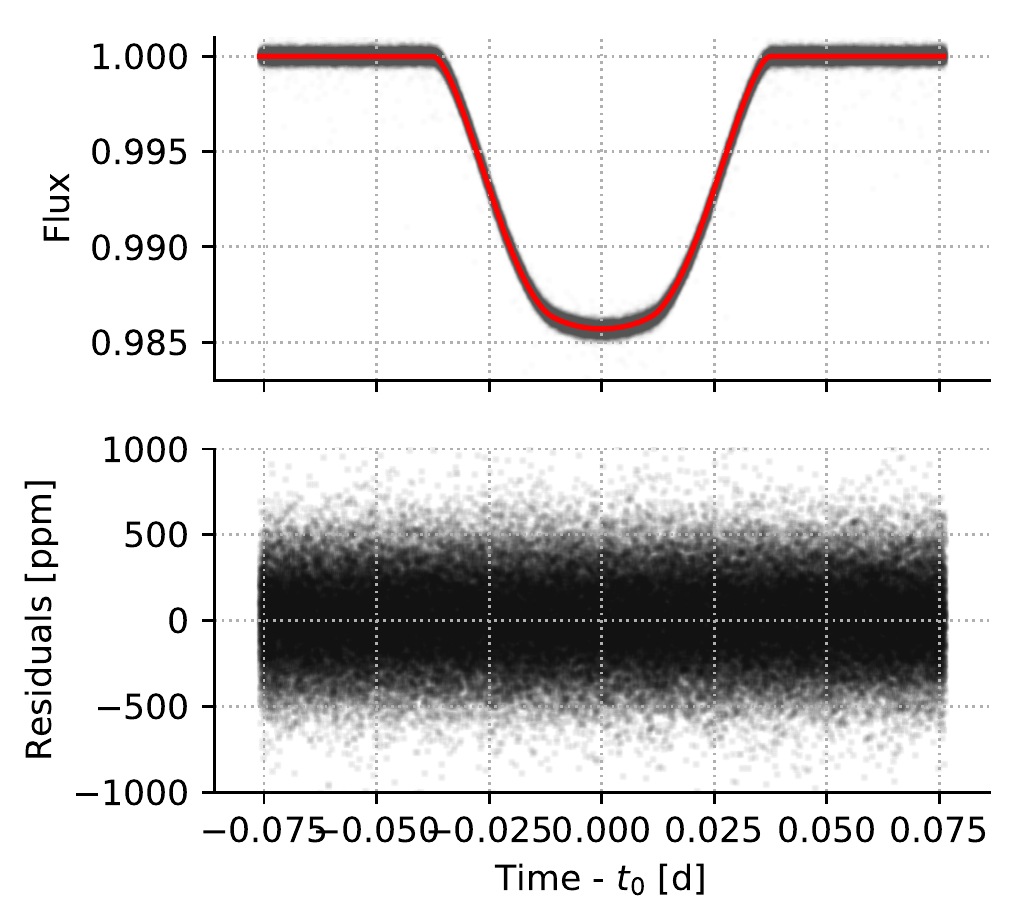}
    \includegraphics[scale=0.75]{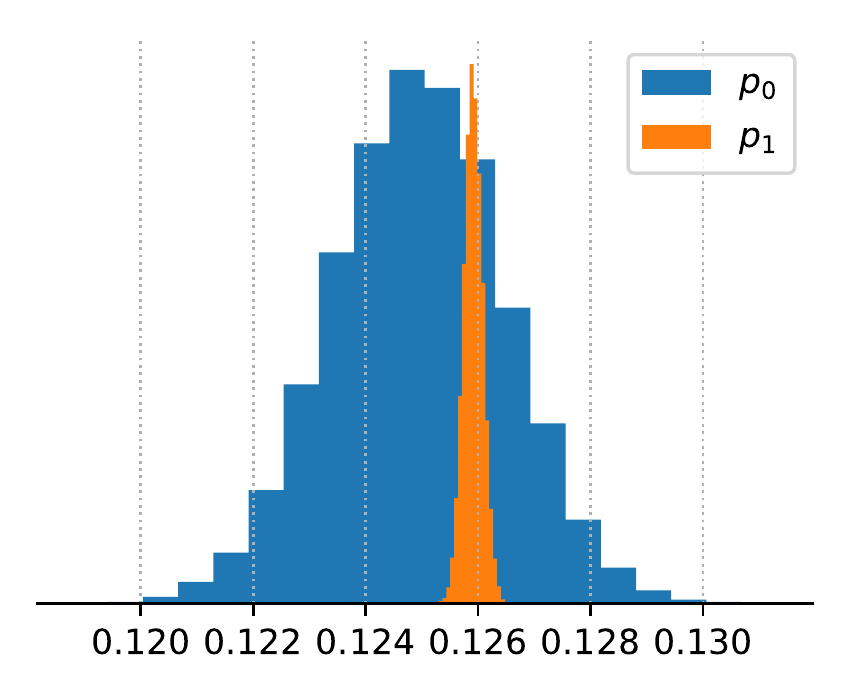}
    \caption{\textsl{Left}: maximum-likelihood transit model for TrES-2 b (red) compared with \kepler short cadence observations (black). \textsl{Right}: posterior distributions verifying that $p_0 \approx p_1$. In other words, we do not detect large-scale surface brightness variations on TrES-2.}
    \label{fig:tres2}
\end{figure*}

TrES-2 is a G0V star with $T_\mathrm{eff} =5960 \pm 100$ K and $\log g = 4.4 \pm 0.2$  \citep{ODonovan2006, Sozzetti2007}. Asteroseismology indicates $M = 0.94 \pm 0.05$ M$_\odot$ and $R = 0.95 \pm 0.02$ R$_\odot$, and the stellar age is $5.8 \pm 2.2$ Gyr \citep{Barclay2012}. The Rossiter-McLaughlin effect shows the orbit of its hot-Jupiter is prograde \citep{Winn2008}. TrES-2 is an especially enticing target because its high impact parameter, $b=0.84$, provides us with long ingress and egress durations and therefore tighter constraints on $p_0$. 

We fit the transit light curve in Figure~\ref{fig:tres2} for $p_0, p_1, i_o, a/R_\star$, holding the quadratic limb darkening parameters fixed at $u_1,u_2 =0.39256, 0.29064$ \citep{Magic2015}. The posterior distributions of $p_0$ and $p_1$ are overlapping, indicating that the planet radius inferred from the transit depth is consistent with the radius inferred from the ingress and egress durations (see Figure~\ref{fig:tres2}). This measurement aligns with our expectation for a middle-aged, solar-like star such as TrES-2, since sunspots typically cover $\sim 0.03\%$ of the solar surface \citep{Howard1984}. We note that the quadratic limb-darkening parameters of \citep{Magic2015} appear to be sufficient for describing the transit of this G0 star even at extremely high S/N.   Table \ref{tab:results}
reports the results from this analysis.

\subsubsection{HAT-P-7 (Kepler-2)}

\begin{figure*}
    \centering
    \includegraphics[scale=0.75]{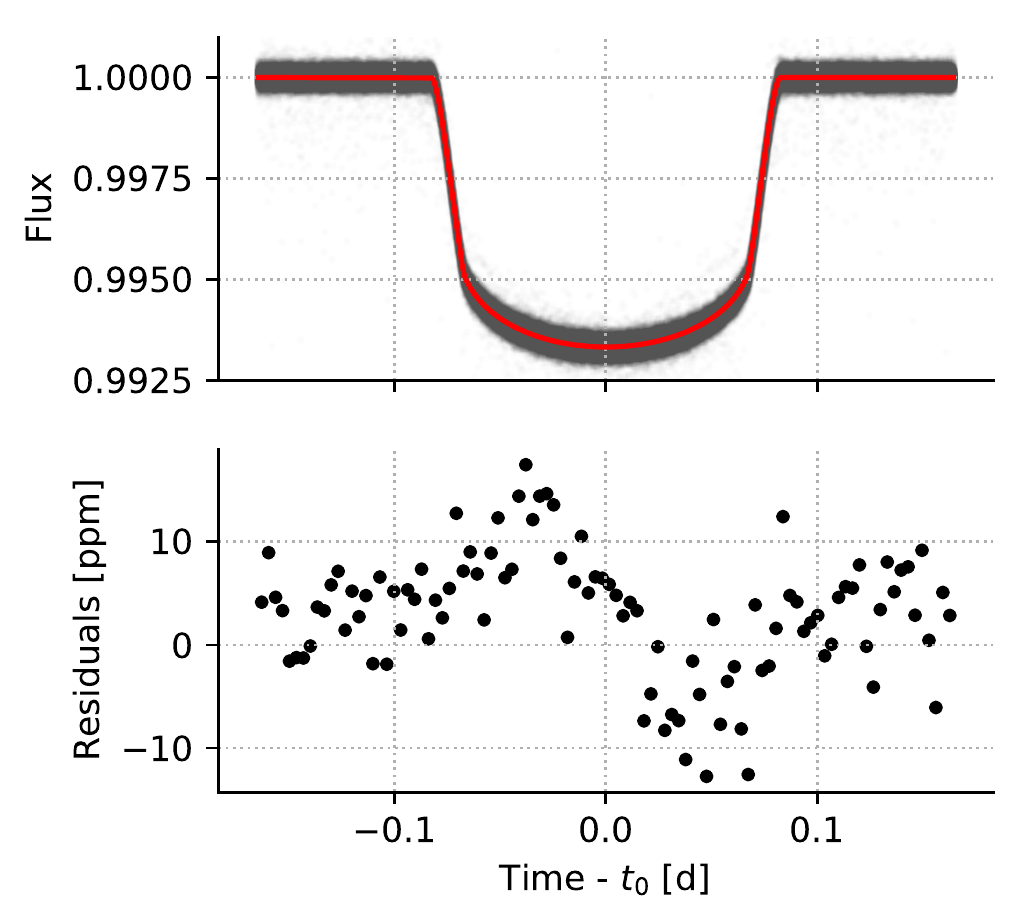}
    \includegraphics[scale=0.75]{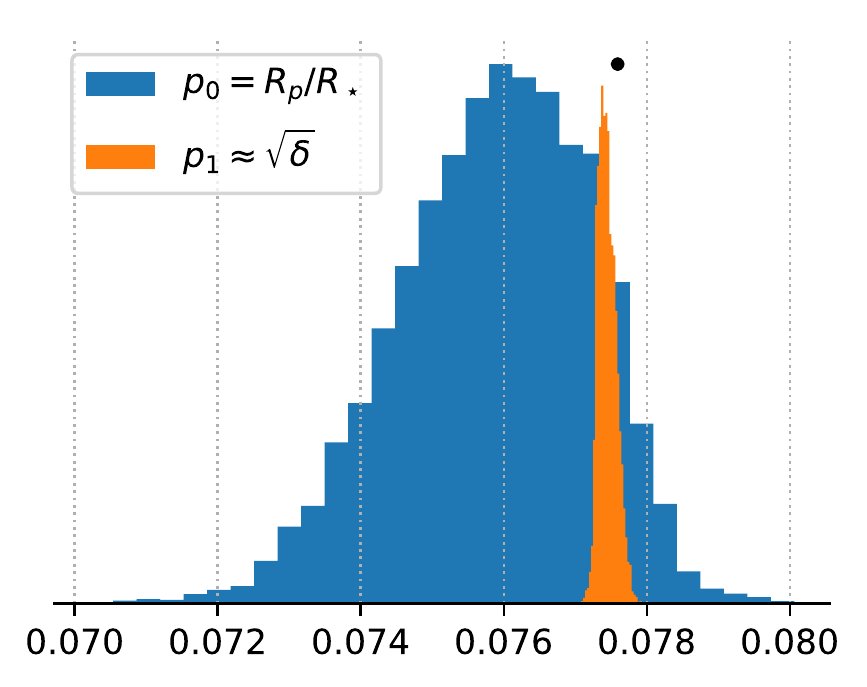}
    \caption{\textsl{Left}: maximum-likelihood transit model for HAT-P-7 b (red) compared with \kepler short cadence observations (black), and binned residuals which show evidence for gravity darkening. \textsl{Right}: posterior distributions for $p_0$ and $p_1$. Note that $p_0 \approx p_1$, indicating that the stellar intensity in the transit chord is generally representative of the star on the whole, despite gravity darkening known to affect this star.}
    \label{fig:hat7}
\end{figure*}

HAT-P-7 is a slightly evolved F6 star, orbited by a highly misaligned hot-Jupiter \citep{pal2008,Winn2009}. Asteroseismology indicates that the star is likely in a pole-on configuration \citep{Lund2014,Benomar2014}. Several authors have noted that there is a $\sim 20$ ppm positive residual bump in the transit light curve between ingress and mid-transit, which may be attributed to stellar gravity darkening \citep{VanEylen2012,Morris2013,Masuda2015}.

We fit the phase-folded \kepler light curve of HAT-P-7 for $p_0, p_1, a/R_\star, i_o, t_0$ and the four nonlinear limb-darkening parameters. We put a Gaussian prior on $a/R_\star$ with the value inferred from asteroseismology by \citep{Lund2014}, $\left< \rho_\star \right> =  0.266 \pm 0.008$ g cm$^{-3}$ \, or $ a/R_\star= 4.090 \pm 0.044$. The results are shown in Figure~\ref{fig:hat7}. We find that $p_0 \approx p_1$, indicating that the transit chord of the planet is representative of the typical intensity of the star, despite the gravity-darkening signal evident in the transit residuals. This would appear to be a result of the stellar orientation, with its bright pole being occulted between mid-transit and egress, and its dimmer low latitudes being occulted between ingress and mid-transit. The net effect of the gravity darkening on the light curve residuals appears to cancel out since $p_0$ is consistent with $p_1$
 

\subsubsection{HAT-P-11 (Kepler-3)} \label{sec:hat11}

\begin{figure*}
    \centering
    \includegraphics[scale=0.75]{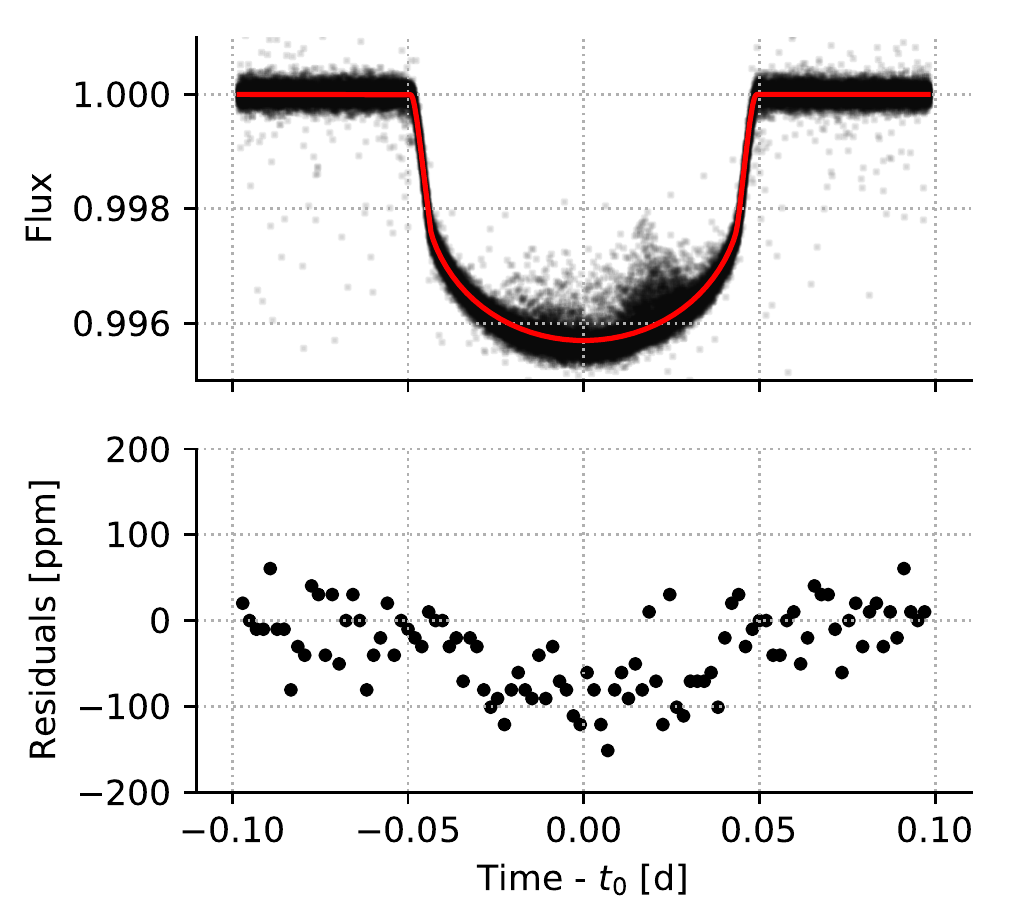}
    \includegraphics[scale=0.75]{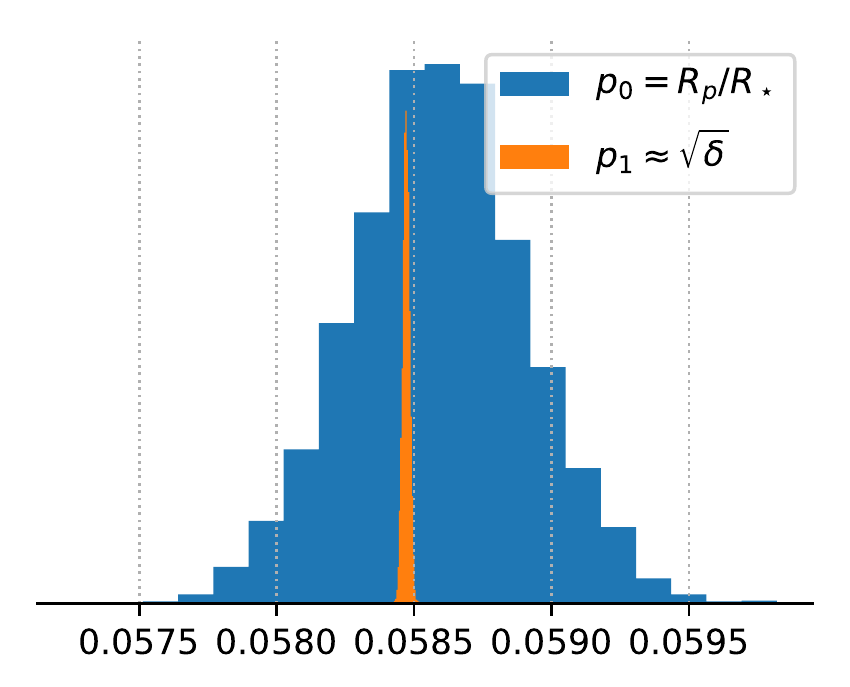}
    \caption{\textsl{Left}: maximum-likelihood transit model for HAT-P-11 b (red) compared with \kepler short cadence observations (black), and binned residuals of the mode within each bin. \textsl{Right}: posterior distributions for $p_0$ and $p_1$. Note that $p_0 \approx p_1$ despite the spots known to be present on HAT-P-11. It would appear that the small spot covering fraction ($f_S \sim 3\%$ from \citealt{Morris2017a}) is insufficient to be measured from the \kepler photometry with this technique.}
    \label{fig:hat11}
\end{figure*}

HAT-P-11 is a benchmark system for studying stellar activity with planetary transits. It is a $0.8 M_\odot$ star with a rotation period similar to the sun ($P_\mathrm{rot} = 29$ d) \citep{bakos2010,Sanchis-Ojeda2011,Southworth2011,Deming2011}. In \citet{Morris2017a,Morris2017b,Morris2018d} we establish qualitative similarities between the Sun and HAT-P-11's starspot distributions and activity cycles. 

We fit the phase-folded \kepler light curve of HAT-P-11 for $p_0, p_1, a/R_\star, i_o, t_0$ and quadratic limb-darkening parameters. We place a Gaussian prior on $a/R_\star$ with the value inferred from asteroseismology by \citet{Christensen-Dalsgaard2010}, $\left< \rho_\star \right> = 2.5127 \pm 0.0009 $ g cm$^{-3} \, \Rightarrow a/R_\star= 14.6950 \pm 0.0017$, and fix the eccentricity and argument of periastron to the measurements of \citet{Winn2010}. The results are shown in Figure~\ref{fig:hat11}. We find $p_0 \approx p_1$, indicating that the spot covering fraction is relatively small. 

It is interesting to compare the spot covering fractions measured in- and out-of-transit for HAT-P-11 from \citet{Morris2017a}. The \kepler light curve shows $\sim3$\% rotational modulation each quarter. If we assume the starspots have contrast similar to sunspots, we can use Equations 9-14 of \citet{Morris2017a} to estimate the spot covering fraction within the transit chord from $p_0$ and $p_1$ by noticing that the depth of transit will be diluted by starspots with contrast $c$ (where $c=1$ means spots with the same intensity as the photosphere and $c=0$ means spots that are perfectly dark), and covering fraction $f_S$,
\begin{equation}
p_1^2 \approx p_0^2 (1 - f_{S,\mathrm{transit}} (1-c)).
\end{equation}
For example, if the transit chord is completely covered ($f_S=1$) by perfectly dark spots ($c=0$), then the transit depth goes to zero. If the transit chord has sun-like spots with $c=0.7$, then we can solve for the spot covering fraction given $p_0$ and $p_1$,
\begin{equation}
f_{S,\mathrm{transit}} \approx \frac{1}{1-c}\left(1 - \left(\frac{p_1}{p_0} \right)^2\right). \label{eqn:fs_transit}
\end{equation}

For HAT-P-11, this yields the spot covering fraction in the transit chord $f_{S,\mathrm{transit}} \approx 0.01 \pm 0.03$. We can compare the in-transit spot coverage to the spot covering fraction that one would estimate from the flux deficit method \citep{Morris2017a}, which estimates the asymmetric component of the spot covering fraction on the whole star via
\begin{equation}
f_{S, \mathrm{asym}} \gtrsim (1 - \min(f))/(1 - c),
\end{equation}
or $f_{S, \mathrm{asym}} \gtrsim 0.03_{-0.01}^{+0.06}$ for HAT-P-11. Detailed spot occultation modeling from \citet{Morris2017a} yields spot covering fractions of 0-10\% within the chord of any given transit, broadly consistent with the small covering fraction measured with the self-contamination technique.  

We note that due to the significant obliquity of the orbit of the planet relative to the spin
of the star,
the time-averaged surface brightness contrast of the transit chord may match the stellar disk
as the spots traverse about the same fraction of the stellar disk as they do of the transit chord.  Consequently, the lack of detection in this case may be hampered by this geometry. 

\subsubsection{Estimating stellar densities in the absence of asteroseismology} \label{sec:huberapprox}

\begin{figure}
    \centering
    \includegraphics[scale=0.8]{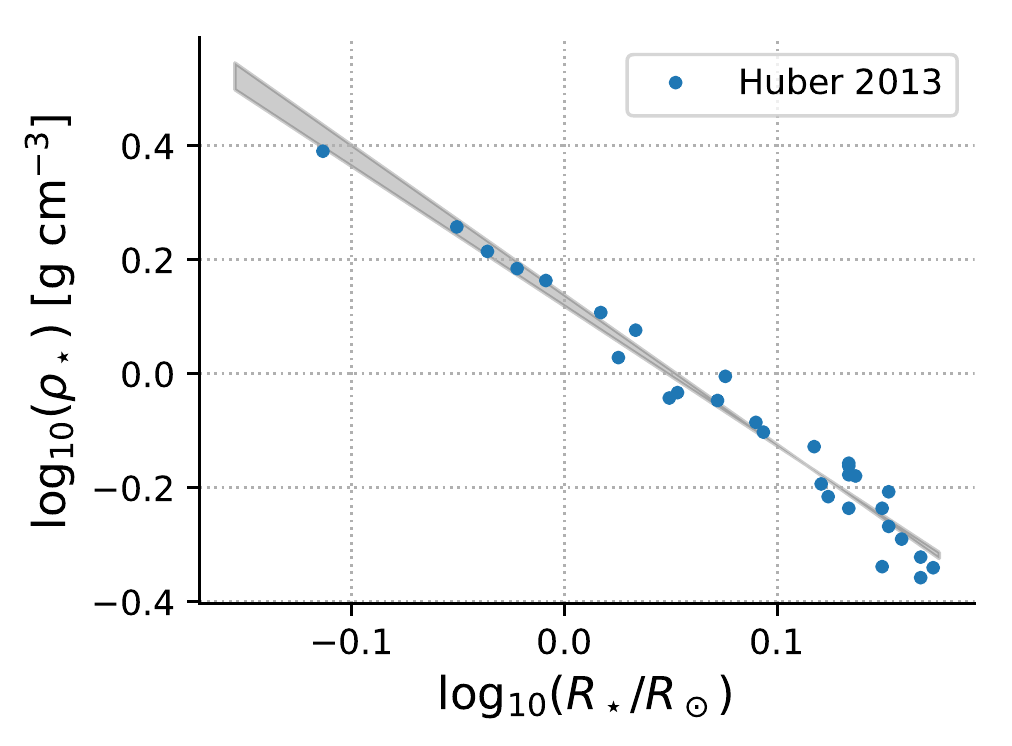}
    \caption{Asteroseimic stellar densities and model stellar radii from \citet{Huber2013} yield constraints on the stellar density for main sequence stars of a given radius. When fitting the following transit light curves, we fix a Gaussian prior on the stellar radii from \citet{Berger2018} via \gaia DR2 observations, and a prior on the stellar density via the relation above and in Equation~\ref{eqn:huber}.}
    \label{fig:huber}
\end{figure}

When a star lacks an asteroseismic density measurement, the three-way degeneracy between $p_0$, the limb-darkening parameters, and the impact parameter (or equivalently $a/R_\star$ or $\rho_\star$) must be broken in order to produce meaningful constraints on $p_0$. For the following three planetary systems, Kepler-17, -39 and -412, no asteroseismic density is available in the literature.

We can estimate the stellar densities for stars without asteroseismic measurements by noting the tight correlation between stellar radius and density found in the asteroseismic sample of \citet{Huber2013} -- see Figure~\ref{fig:huber}. For bright \kepler targets, \citet{Huber2013} used asteroseismology to calculate stellar densities, and isochrone models to predict stellar radii and masses. We fit a power-law to the measured stellar densities as a function of radius, and find 
\begin{equation}
\begin{split}
\log_{10} \left(\rho_\star [\mathrm{g \, cm}^{-3}] \right) = \\ \left( -2.540 \pm 0.087 \right) \log_{10}  \left( R_\star/R_\odot \right) + \left( 0.12 \pm 0.001 \right), \label{eqn:huber}
\end{split}
\end{equation}
with fractional scatter in $\rho_\star$ of 0.083. Then for each star, we use the \gaia DR2 \citep{Gaia2018} stellar radius measurements from \citet{Berger2018} in combination with Equation~\ref{eqn:huber} to predict the stellar density. 

At each step in our Markov chains for Kepler-17, -39 and -412, we sample for the parameters $M_\star$ and $R_\star$, placing a Gaussian prior on $R_\star$ with the value and uncertainty from \gaia \citep{Berger2018} ($R_\star = 1.01R_\sun$, $1.23 R_\sun$, $1.35 R_\sun$ for Kepler-17, -36 and -412, respectively), and placing a Gaussian prior on the stellar density following Equation~\ref{eqn:huber} with fractional uncertainty 0.083 g cm$^{-3}$. 

\subsubsection{Kepler-17} \label{sec:k17}

Kepler-17 is an active G2V star with $T_\mathrm{eff} = 5630 \pm 100$ K and age $3 \pm 1.6$ Gyr, with a hot Jupiter companion that produces starspot occultations in nearly every transit studied by \citet{Desert2011} and \citet{Davenport2015thesis}. Thus Kepler-17's geometry is similar to the toy model in Section~\ref{sec:activelat}, where a planet occults an active latitude with many starspots, diluting the transit depth, and in which the stellar spin
axis is aligned (in this case, to within 15$^\circ$).

\begin{figure*}
    \centering
    \includegraphics[scale=0.75]{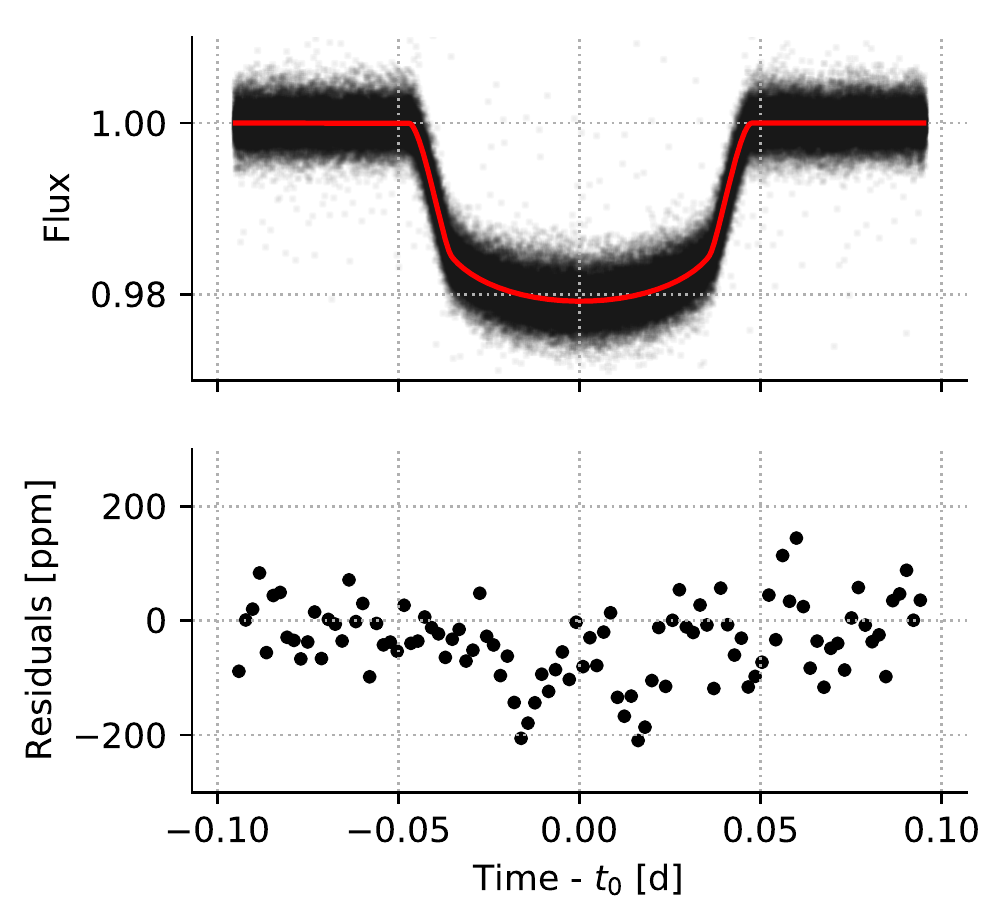}
    \includegraphics[scale=0.75]{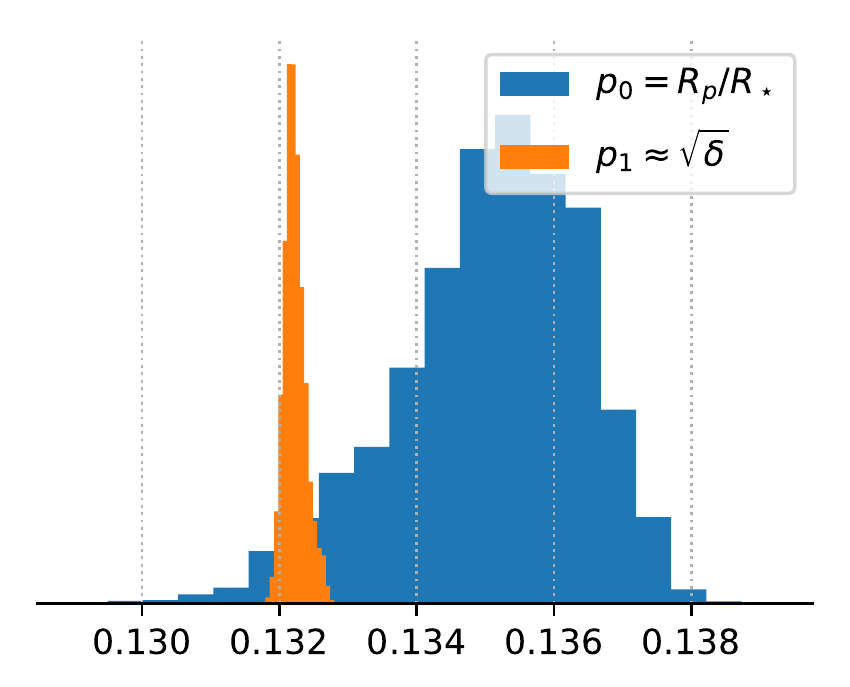}
    \caption{\textsl{Left}: maximum-likelihood transit model for Kepler-17 b (red) compared with \kepler short cadence observations (black). \textsl{Right}: Posterior histograms for $p_0, p_1$. For this system we find weak evidence for $p_0 > p_1$, consistent with an occultation of a dark region of the stellar surface. This is consistent with the detailed analyses of \citet{Desert2011} and \citet{Davenport2015thesis}, which suggest that Kepler-17 occults an active latitude near the stellar equator.
    }
    \label{fig:kepler17_lc}
\end{figure*}

We fit the phase-folded \kepler light curve of Kepler-17 for $p_0, p_1, b, t_0, T_{14}, M_\star, R_\star$, and quadratic limb-darkening parameters. We place priors on $M_\star$ and $R_\star$ as described in Section~\ref{sec:huberapprox}. The results shown in Figure~\ref{fig:kepler17_lc} indicate $p_0 > p_1$ at 93\% confidence ($\sim 2 \sigma$), in agreement with our expectation that starspot occultations dilute the apparent transit depth. We measure $p_0 = R_p/R_\star = 0.1354_{-0.0014}^{+0.0010}$, which is significantly larger than the \citet{Desert2011} estimate ($R_p/R_\star = 0.13031^{+0.00022}_{-0.00018}$).

As we did for HAT-P-11 in Section~\ref{sec:hat11}, we can compare the spot covering fractions measured in- and out-of-transit for Kepler-17. The \kepler light curve shows $\sim4$\% rotational modulation each quarter. If we assume the starspots have contrast similar to sunspots, as in \citet{Davenport2015thesis}, we can use Equation~\ref{eqn:fs_transit} to estimate the spot covering fraction within the transit chord and find $f_{S,\mathrm{transit}} \approx 0.27_{-0.18}^{+0.12}$. We can compare the in-transit spot coverage to the spot covering fraction that one would estimate from the flux deficit method, $f_{S, \mathrm{asym}} \gtrsim 0.13$. These two measurements are consistent, and perhaps suggest that the transit chord may be more spotted than the mean photosphere, consistent with the apparent equatorial active latitude deduced from detailed spot modeling by \citet{Davenport2015thesis}.

\subsubsection{Kepler-39} \label{sec:k39}

\begin{figure*}
    \centering
    \includegraphics[scale=0.75]{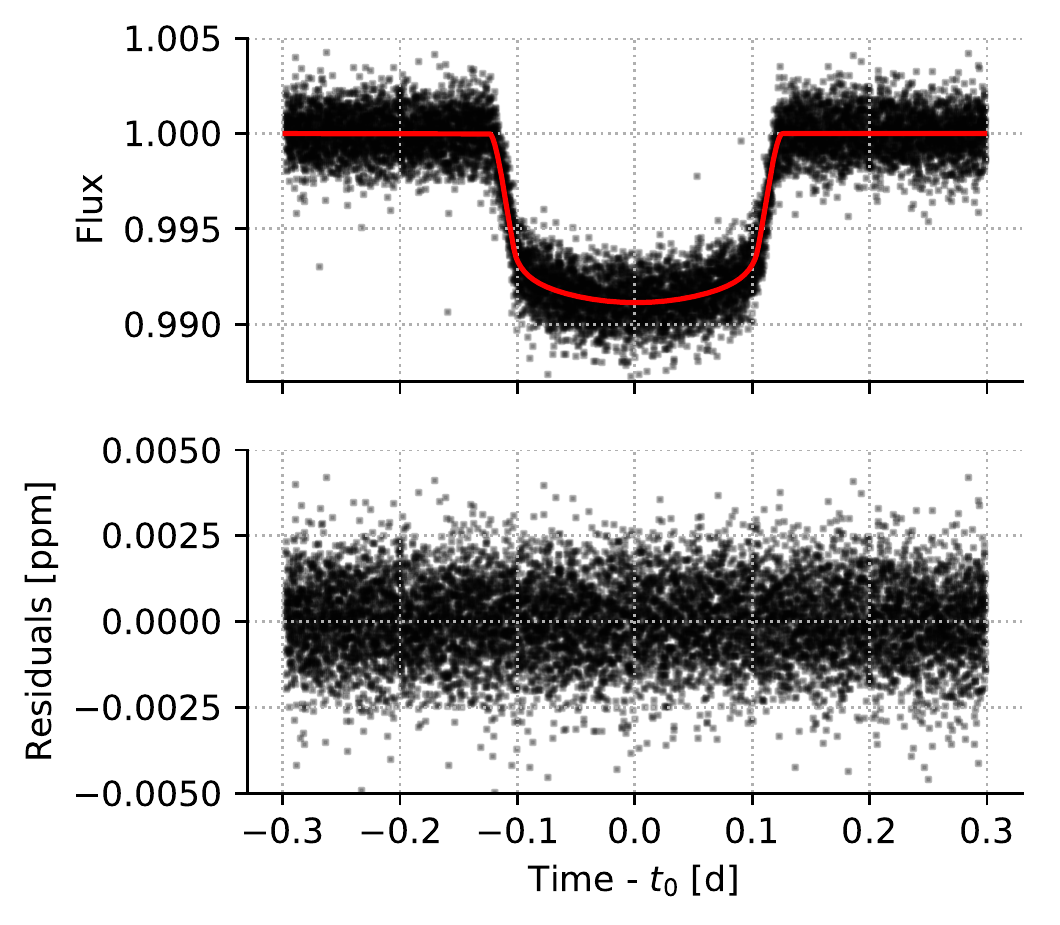}
    \includegraphics[scale=0.75]{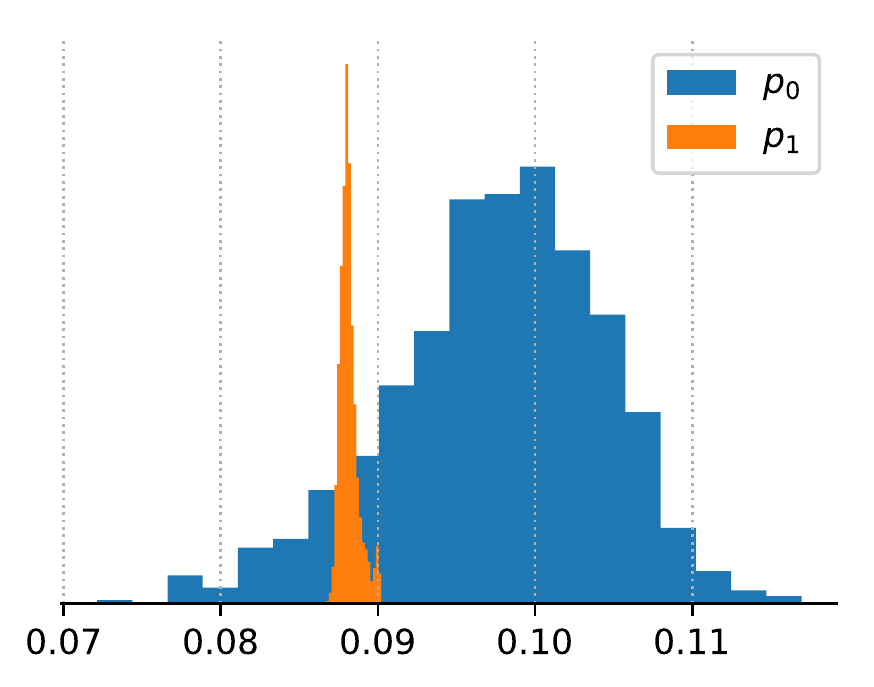}
    \caption{\textsl{Left}: maximum-likelihood transit model for Kepler-39 b (red) compared with \kepler short cadence observations (black). \textsl{Right}: Posterior histograms for $p_0, p_1$. For this system we detect weak evidence for $p_0 > p_1$, consistent with either (1) an occultation of a dark region of the stellar surface; or (2) oblateness of the planetary companion $f = {0.25}^{+0.13}_{-0.06}$. The out-of-transit rotational modulation due to starspots confirms that there are starspots present, see Figure~\ref{fig:rotation_modulation}, while others in the literature interpret the signal as planetary oblateness \citep{Zhu2014}.}
    \label{fig:kepler39_lc}
\end{figure*}

Kepler-39 is a young F7V host star with $T_\mathrm{eff} = 6350 \pm 100$ K, and age $0.7^{+0.9}_{-0.3}$ Gyr \citep{Mamajek2008, Bonomo2015}. \citet{Zhu2014} claim its 20 $M_J$ companion has nonzero oblateness $f = 0.22^{+0.11}_{-0.11}$.

We fit the phase-folded \kepler light curve of Kepler-39 for $p_0, p_1, b, t_0, T_{14}, M_\star, R_\star$, and quadratic limb-darkening parameters. We place priors on $M_\star$ and $R_\star$ as described in Section~\ref{sec:huberapprox}. The results are shown in Figure~\ref{fig:kepler39_lc}. \citet{Zhu2014} measure $R_p/R_\star = 0.0889 \pm 0.0006$ with their oblate planet model, consistent with our maximum likelihood $p_0 = R_p/R_\star = 0.090 \pm 0.010$. We find that $p_0 > p_1$ with 90\% confidence, suggesting weak evidence for either starspot occultations in the transit light curve, or the degenerate signal of planetary oblateness. We find $f = {0.25}^{+0.13}_{-0.06}$, consistent with the oblateness reported by \citet{Zhu2014}.

\begin{figure*}
    \centering
    \includegraphics[scale=0.6]{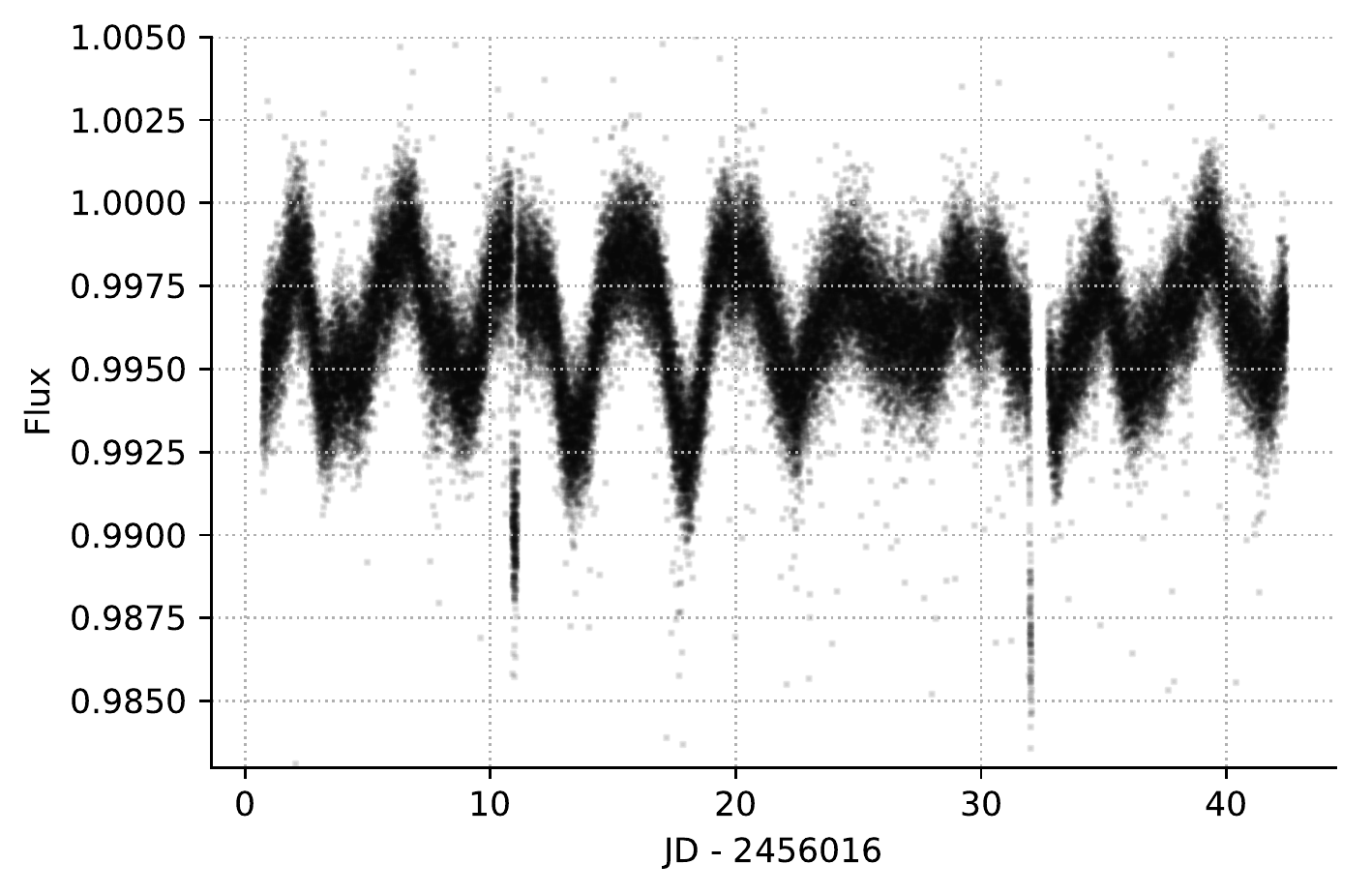}
    \includegraphics[scale=0.6]{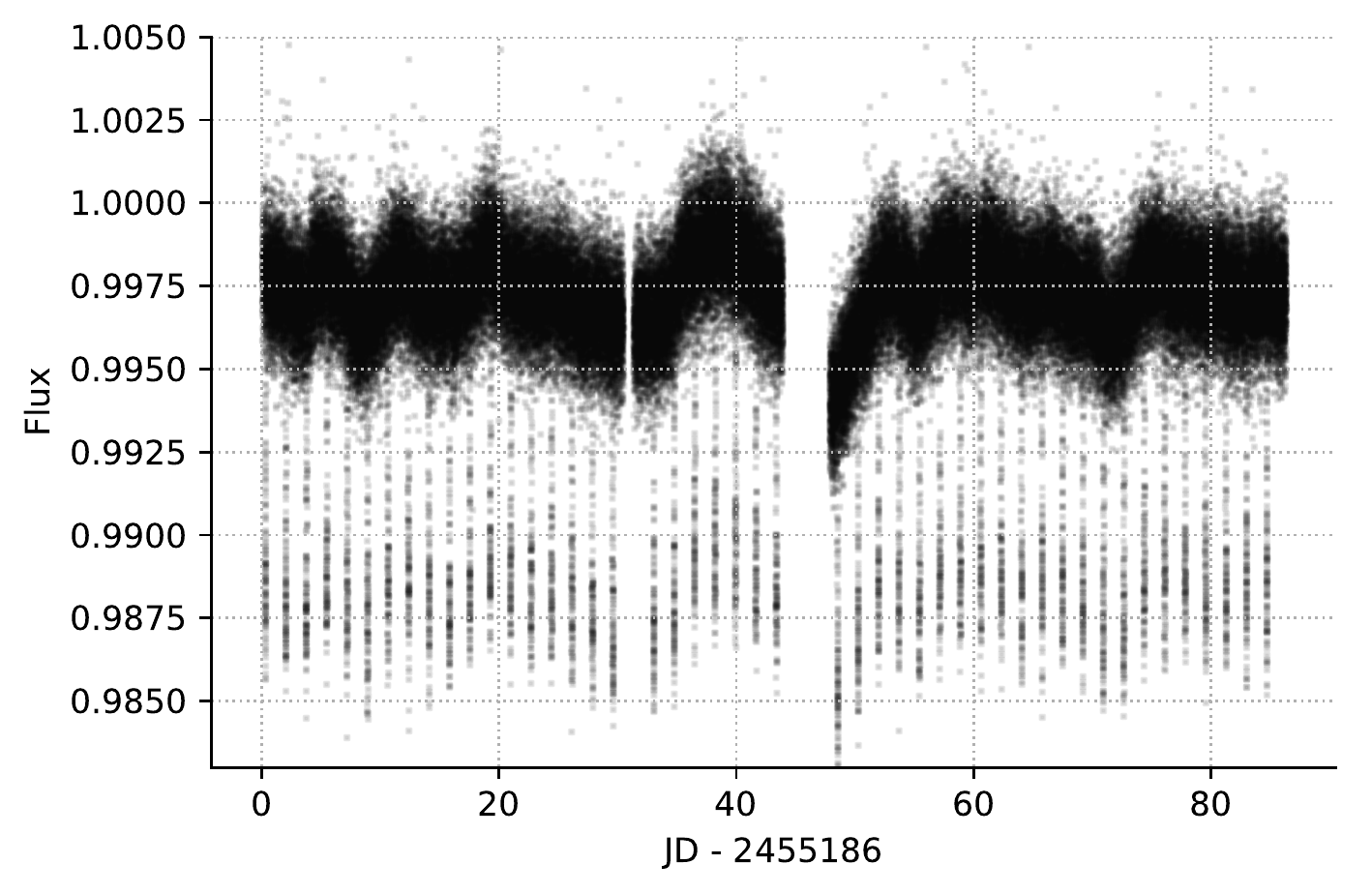}
    \caption{Rotational modulation of Kepler-39 (left) and Kepler-412 (right) showing evidence for starspot coverage, despite the lack of starspot crossings in the transit photometry.}
    \label{fig:rotation_modulation}
\end{figure*}

However, we caution that the \kepler light curve shows rotational modulation of the star consistent with starspot coverage (see Figure~\ref{fig:rotation_modulation}), albeit several times smaller than for Kepler-17. We can estimate the minimum spot covering fraction from the amplitude of the rotational modulation, which gives us a constraint on the asymmetry of the spot distribution as the star rotates, $f_{S,\rm min} \geq (1 - \min{\textrm{flux}}) / (1-c)$ where $c=0.7$ is the spot contrast for sunspots \citep{Solanki2003}. Thus assuming spots with the contrast of sunspots, $f_{S,\rm min} \geq 2\%$ -- a factor of $\gtrsim 100$ greater than the typical sunspot area coverage, comparable to the spot coverage of HAT-P-11 \citep{Morris2017a}. It is possible that the marginal detection of $p_0 > p_1$ is attributable to either or both planetary oblateness and occulted starspots.


\subsubsection{Kepler-412}

\begin{figure*}
    \centering
    \includegraphics[scale=0.75]{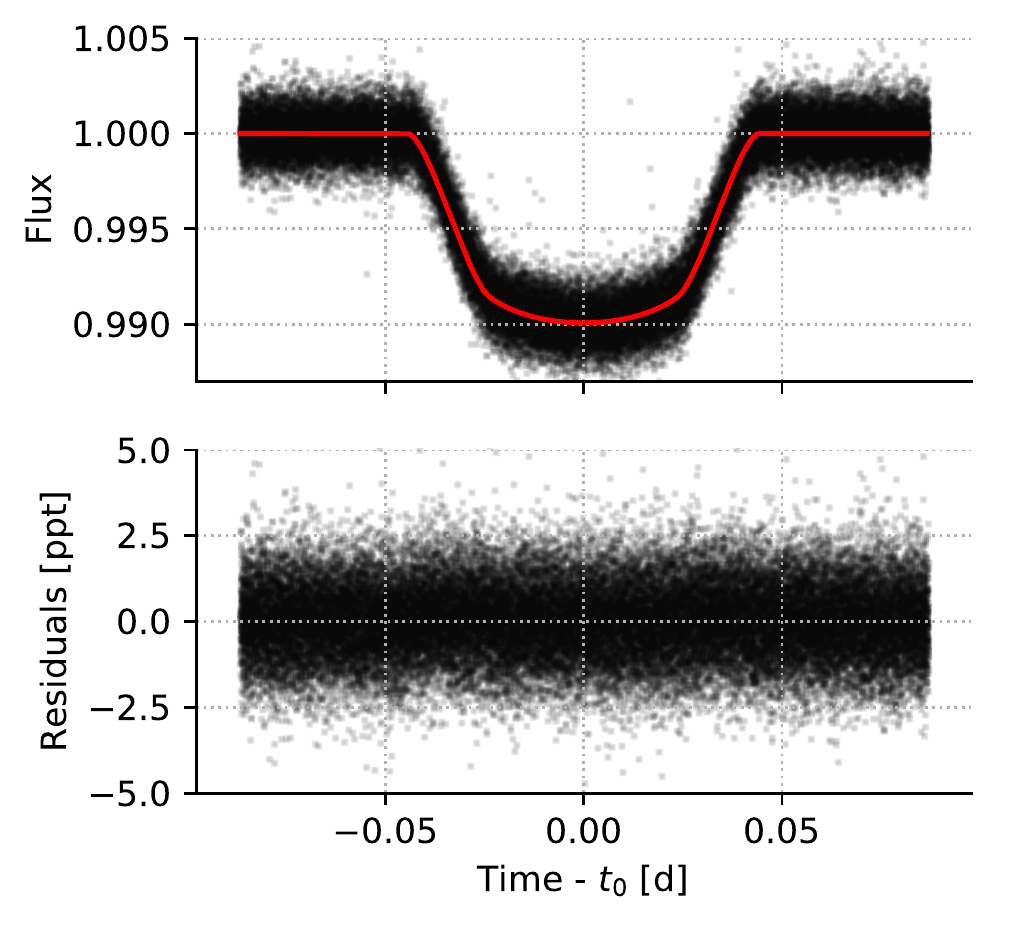}
    \includegraphics[scale=0.75]{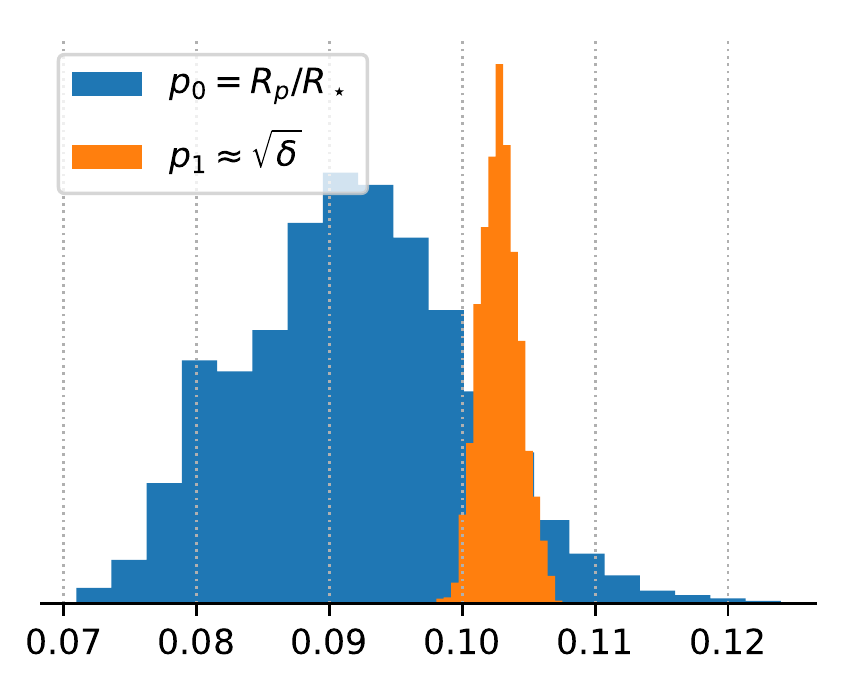}
    \caption{\textsl{Left}: maximum-likelihood transit model for Kepler-412 b (red) compared with \kepler short cadence observations (black). \textsl{Right}: Posterior distributions for $p_0,p_1$ and $R_p/R_\star$ (equivalent to fixing $p_0 = p_1$). There is insignificant evidence for $p_0 < p_1$. See Figure~\ref{fig:rotation_modulation} for the out-of-transit rotational modulation of Kepler-412.}
    \label{fig:kepler412}
\end{figure*}

Kepler-412 is a G3V host to a hot Jupiter with $M_p = 0.939 \pm 0.085 M_J$, and an apparently inflated radius of $1.325 \pm 0.043 R_J$ \citep{Deleuil2014}. 

We fit the phase-folded \kepler light curve of Kepler-412 for $p_0, p_1, b, t_0, T_{14}, M_\star, R_\star$, and quadratic limb-darkening parameters. We place priors on $M_\star$ and $R_\star$ as described in Section~\ref{sec:huberapprox}. The results are shown in Figure~\ref{fig:kepler412}; we find $p_0 = R_p/R_\star = 0.0926_{-0.0089}^{+0.0095}$ corresponding to $R_p = 1.16 \pm 0.11 R_J$ -- consistent with the literature value ($R_p/R_\star = 0.1058 \pm 0.0023$, \citealt{Deleuil2014}). We measure $p_0 \approx p_1$. 

As with Kepler-39 in Section~\ref{sec:k39}, $p_0 \lesssim p_1$ could be interpreted as insignificant evidence ($\sim 1 \sigma$) for dark starspots on the stellar photosphere outside of the transit chord;  however, without the ability to further refine the uncertainty on $p_0$, we cannot strengthen this claim. The \kepler light curve shows small variations in the stellar intensity that may be rotational modulation due to unocculted starspots (see Figure~\ref{fig:rotation_modulation}).

\subsection{K2 Observations of TRAPPIST-1}

TRAPPIST-1 is an M8V host to seven Earth-sized planets \citep{Gillon2016,Gillon2017,Luger2017,Delrez2018}. Its spots seem to evolve on a timescale similar to the apparent rotation period \citep{Roettenbacher2017}, the spot and faculae covering fractions might be quite high \citep{Rackham2018,Zhang2018}, and bright spots might be required to explain the apparent rotational modulation of TRAPPIST-1 \citep{Morris2018c}. Here we analyze the K2/EVEREST short-cadence TRAPPIST-1 light curves of the two innermost TRAPPIST-1 planets for evidence of photospheric inhomogeneities \citep{Luger2016,luger2017everest}.

\begin{figure*}
    \centering
    \includegraphics[scale=0.75]{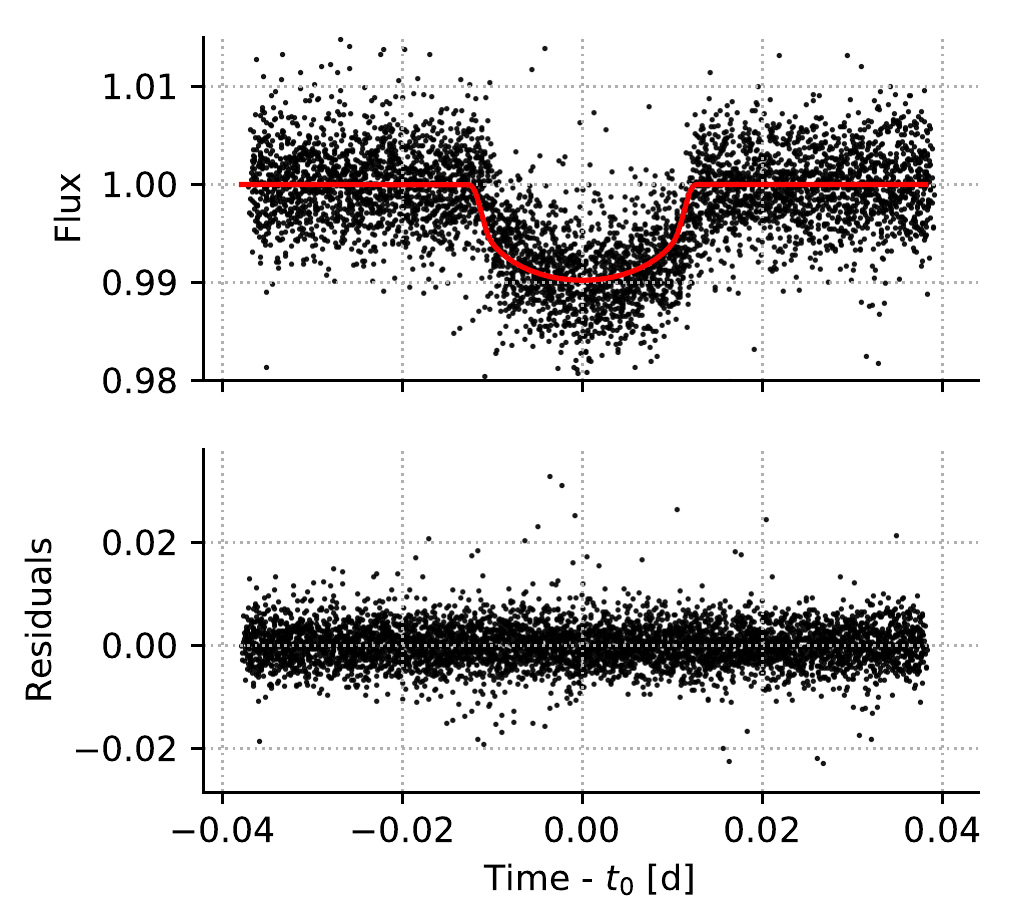}
    \includegraphics[scale=0.75]{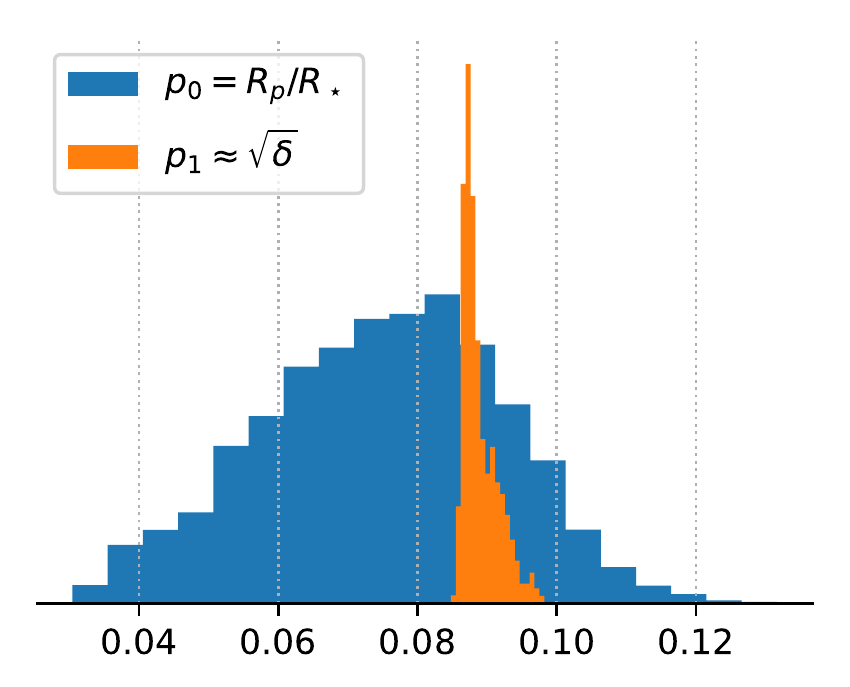}
    \caption{\textsl{Left}: maximum-likelihood transit model for TRAPPIST-1 b (red) compared with \kepler/K2 short cadence observations (black). \textsl{Right}: posterior distributions for $p_0$ and $p_1$. $p_0$ is consistent with $p_1$, suggesting the occulted stellar surface is generally similar in brightness to the rest of the surface. }
    \label{fig:t1b}
\end{figure*}

\begin{figure*}
    \centering
    \includegraphics[scale=0.75]{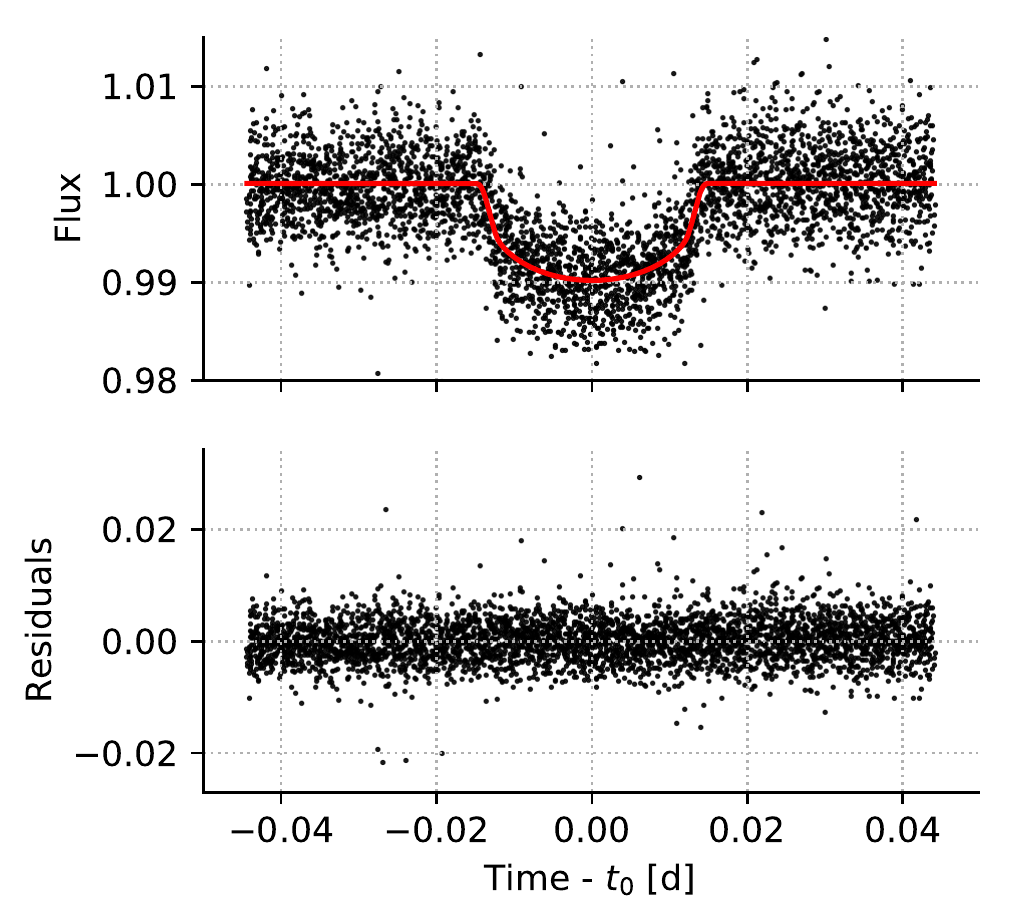}
    \includegraphics[scale=0.75]{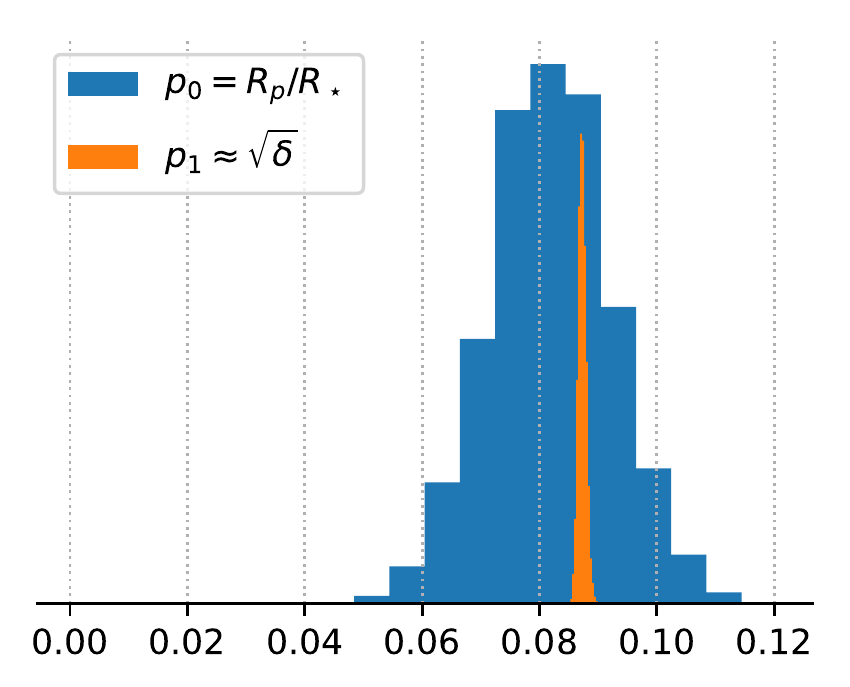}
    \caption{\textsl{Left}: maximum-likelihood transit model for TRAPPIST-1 c (red) compared with \kepler/K2 short cadence observations (black). \textsl{Right}: posterior distributions for $p_0$ and $p_1$. See Figure~\ref{fig:t1b} for further interpretation.}
    \label{fig:t1c}
\end{figure*}

We fit the transit light curves of TRAPPIST-1 b and c for $p_0, p_1, T_{14}$ and the quadratic limb darkening coefficients $u_1$ and $u_2$, see Figures \ref{fig:t1b} and \ref{fig:t1c}. We place Gaussian priors on the impact parameter based on the joint light curve analysis by \citep{Luger2017}, with impact parameters for planets b and c: $b_b = {0.093}_{-0.015}^{+0.011}$ and
$b_c = {0.128}_{-0.020}^{+0.015}$. 

We find $p_0 \approx p_1$ for both planets, indicating insufficient evidence for a difference in spot coverage inside compared to outside of the overlapping transit chords of planets b and c. Combined with the lack of observed spot occultations, the indication that the star may not be highly spotted is somewhat at odds with the spot covering fraction estimates from \citet{Rackham2018, Zhang2018}, who find a potentially large spot covering fraction, $f_S = 8^{+18}_{-7} \%$.  
As best we can tell from Doppler imaging, fully-convective stars are likely highly spotted, and their spots may be randomly distributed, arranged into active latitudes, or concentrated at the poles \citep{Barnes2001,Donati2003,Morin2008}. If we continue to measure $p_0 \approx p_1$ with further transit observations of TRAPPIST-1, that may suggest that active regions on the star are small,  low contrast, and/or uniformly distributed. It will be especially interesting to search for differences between $p_0$ and $p_1$ for the outer planets, which span a range of impact parameters, and thus a range of stellar latitudes. Upcoming \spitzer and JWST observations may improve this measurement, as the degeneracy between $p_0$ and the limb darkening is substantially diminished in the infrared, although the spot contrast is also diminished at longer wavelengths.

\subsection{\spitzer Light Curves} \label{sec:spitzer}

We analyze two \spitzer light curves of interest. One is the transit of HD 80606 b, which is remarkable -- among other reasons -- for having a twelve hour transit observed at a very short cadence, providing us with exceptionally well-sampled transit ingress and egress. The other is GJ 1214 -- an M dwarf host to a transiting super-Earth.

\subsubsection{HD 80606}

\begin{figure*}
    \centering
    \includegraphics[scale=0.75]{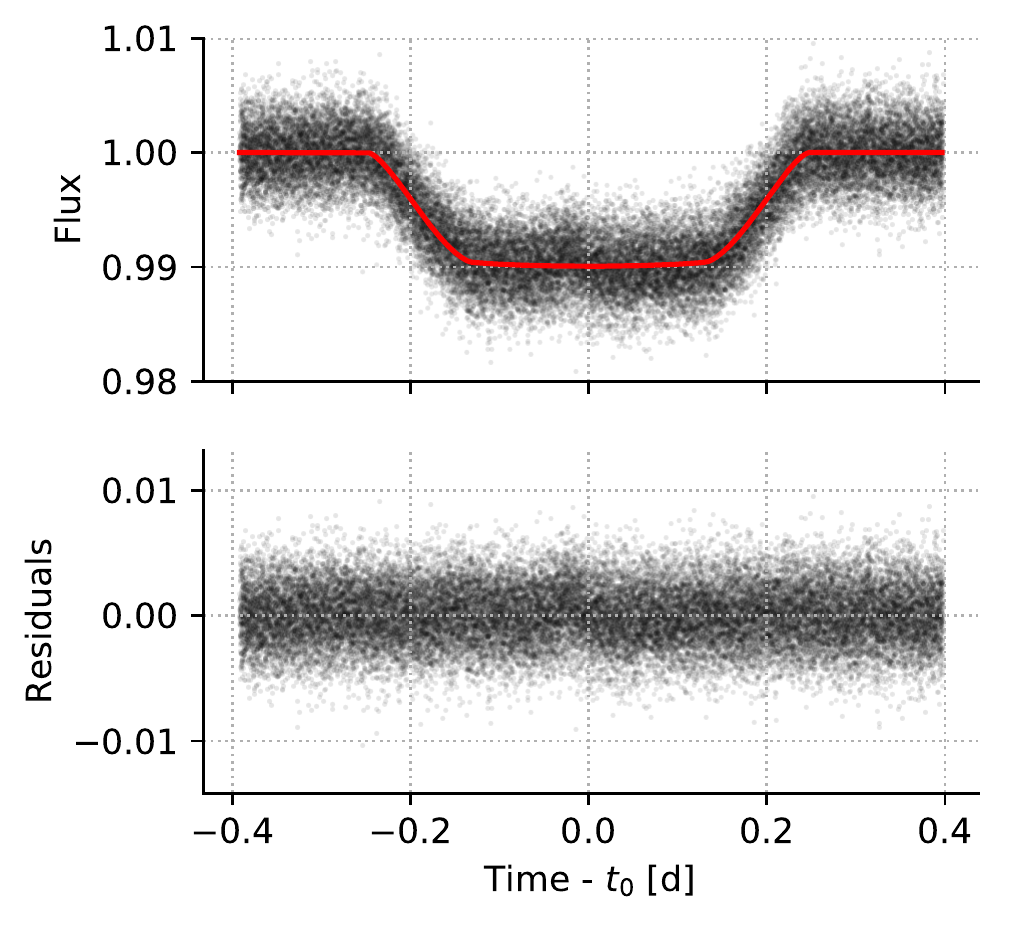}
    \includegraphics[scale=0.75]{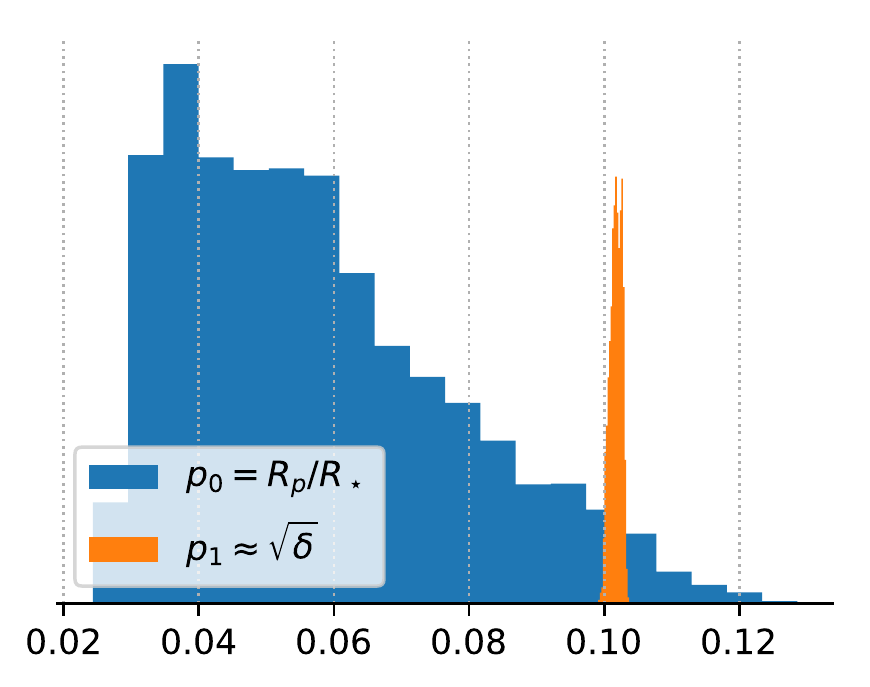}
    \includegraphics[scale=0.6]{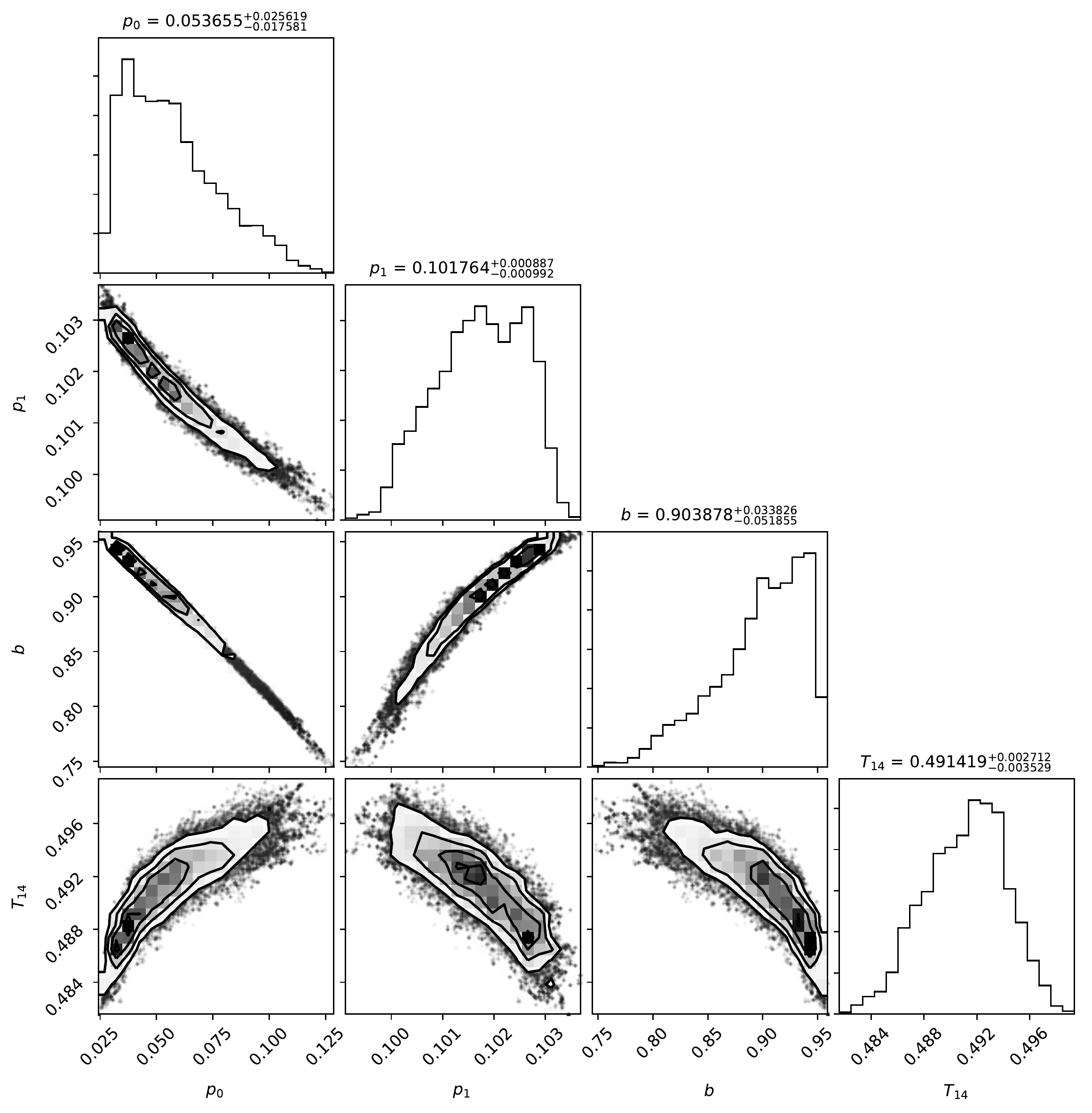}
    \caption{\textsl{Upper left}: maximum-likelihood transit model for HD 80606 b (red) compared with the \spitzer observations at 4.5 $\mu$m from \citet{Hebrard2010} (black), and binned residuals. \textsl{Upper right}: \textsl{Right}: Posterior distributions for $p_0,p_1$ and $R_p/R_\star$ (equivalent to fixing $p_0 = p_1$). The modest difference in $p_0$ and $p_1$ may indicate that there is stellar activity and the transit chord is an unusually bright portion of the stellar surface (see Section~\ref{sec:brightlat}). \textsl{Lower}: Posterior distributions for all parameters in the fit to the transit light curve of HD 80606 b.}
    \label{fig:hd80606}
\end{figure*}

HD 80606/HD 80607 is a wide, solar-type binary system with a massive Jovian planet orbiting the brighter component in an extremely eccentric ($e = 0.93$), misaligned orbit \citep{Naef2001, Moutou2009, Winn2009b}. We re-analyze warm \spitzer photometry of the $\sim 12$ hour-long transit observed at 4.5 $\mu$m by \citet{Hebrard2010} (see Figure~\ref{fig:hd80606}).

We fit the \spitzer transit light curve of HD 80606 b for $p_0, p_1, b, t_0$ and $T_{14}$, fixing the quadratic limb-darkening parameters $u_1, u_2=0.0866, 0.1071$ \citep{Claret2013}. We find $p_0 < p_1$ with 96\% confidence, suggesting that either the planet may be occulting a relatively bright region of the stellar surface, or that we have to place a constraint upon the impact parameter.  The transit light curve of HD 80606 shows evidence of at least one starspot occultation; perhaps  unocculted starpsots are responsible for the tension between $p_0$ and $p_1$ (see Figure~\ref{fig:hd80606}). 

The maximum likelihood $p_0 = R_p/R_\star =  0.053^{+0.025}_{-0.017}$ in combination with the stellar radius from \citet{Hebrard2010} of $1.007 \pm 0.024 R_\odot$ yields the planetary radius $R_p = 0.51_{-0.16}^{+0.26} R_J$, $2\sigma$ smaller than proposed by \citet{Hebrard2010} ($R= 0.981 R_J$);  they also measure a smaller impact parameter, $b=0.808 \pm 0.007$, than we measure, which accounts for this discrepancy.

The density of the star inferred by \citet{Hebrard2010} is $\rho_* = 1.39\pm 0.07$ g/cc, which is consistent
with the value we infer from the \citet{Huber2013} relation that we derived above of $\rho_{*,Huber} = 1.29 \pm 0.11$ g/cc.
We conclude that if we had placed this density constraint as a prior on the stellar density, we would have
obtained a smaller impact parameter, and hence a larger value of $p_0$ (see Figure \ref{fig:corners}) consistent
with \citet{Hebrard2010}.

\subsubsection{GJ 1214}

GJ 1214 is an M4.5 host to a transiting super-Earth \citep{Charbonneau2009}. The stellar activity of GJ 1214 has been studied extensively for its effect on the transmission spectrum of this potentially water-rich world \citep{Fraine2013}. Rotational modulation of the stellar flux is very limited, suggesting that either the star has either few or small spots, or more likely that the spots are arranged in a nearly-axisymmetric distribution \citep{Berta2011,Narita2013}. We re-analyze warm \spitzer 4.5 $\mu$m photometry of 13 transits of GJ 1214 b by \citet{Gillon2014} to verify that the spot distribution inside the transit chord is similar to the rest of the stellar surface. 

\begin{figure*}
    \centering
    \includegraphics[scale=0.75]{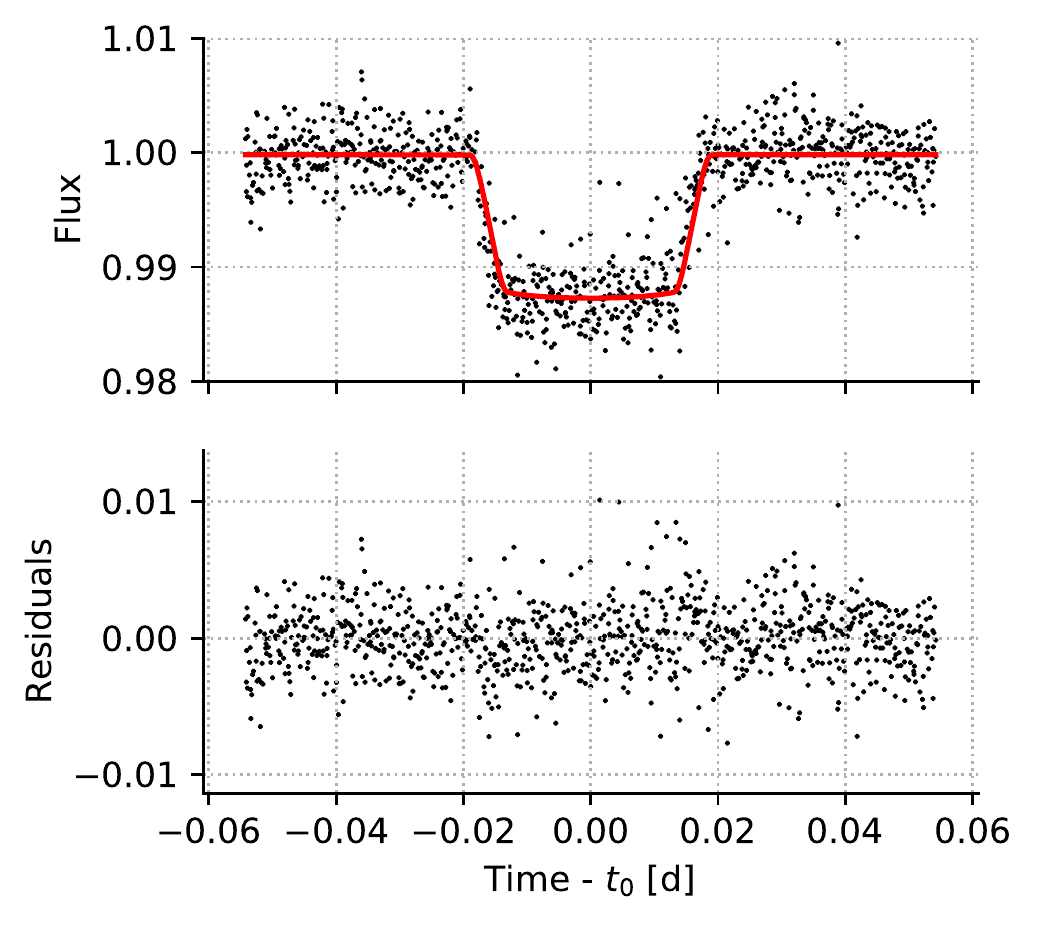}
    \includegraphics[scale=0.75]{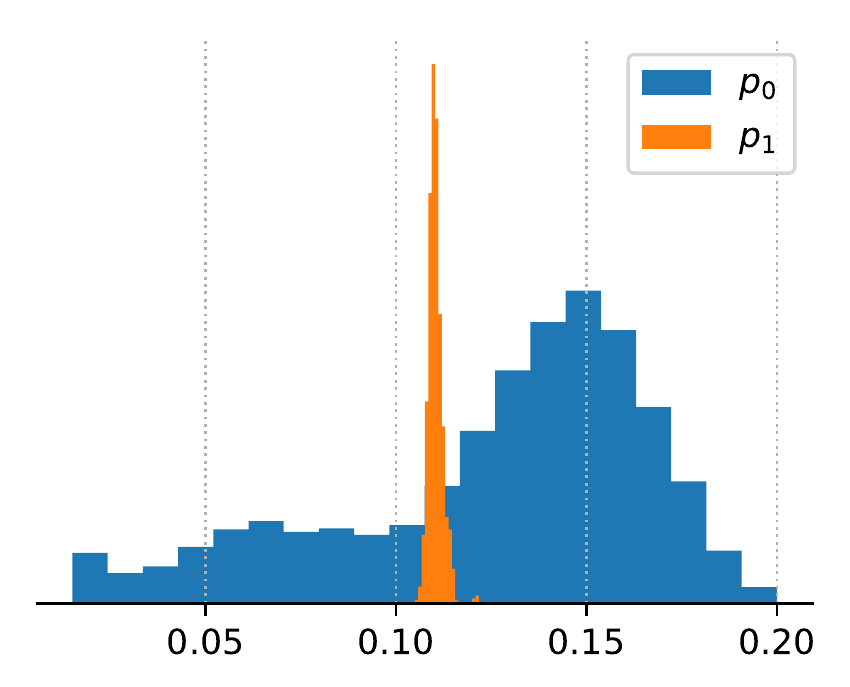}
    \caption{\textsl{Left}: maximum-likelihood transit model for GJ 1214 b (red) compared with the \spitzer observations at 4.5 $\mu$m from \citet{Gillon2014} (black). \textsl{Right}: posterior distributions for $p_0$ and $p_1$. Consistent values for both parameters indicates that the intensity distribution in the transit chord is consistent with the rest of the stellar surface.}
    \label{fig:gj1214}
\end{figure*}

We fit the \spitzer transit light curve of GJ 1214 for $p_0, p_1, b, T_{14}$ and fit for the quadratic limb darkening parameters $q_1$ and $q_2$ \citep{Kipping2013}, see Figure \ref{fig:gj1214}. The resulting constraint on $p_0$ is very weak and is consistent with $p_1$, suggesting an isotropic spot distribution, consistent with the conclusions of \citet{Berta2011} and \citet{Narita2013}. 

\begin{table*}
\centering
\begin{tabular}{l|ccccc}
Target & $p_0=R_p/R_\star$ & $p_1\approx\sqrt{\delta}$ & $\epsilon$ & Condition & CI \\ \hline
TrES-2 & ${0.1250}_{-0.0014}^{+0.0015}$ & ${0.12590}_{-0.00017}^{+0.000168}$ & ${-0.013}_{-0.027}^{+0.027}$ & $p_0 < p_1$ & 71\% \\
HAT-P-7 & ${0.0758}_{-0.0016}^{+0.0015}$ & ${0.07744}_{-0.00013}^{+0.00015}$ & ${-0.042}_{-0.050}^{+0.044}$ & $p_0 < p_1$ & 85\% \\
HAT-P-11 & ${0.05859}_{-0.00031}^{+0.00033}$ & ${0.058471}_{-0.000012}^{+0.000012}$ & ${0.004}_{-0.010}^{+0.011}$ & $p_0 > p_1$ & 66\% \\
Kepler-17 & ${0.1348}_{-0.0017}^{+0.0013}$ & ${0.13224}_{-0.00015}^{+0.00024}$ & ${0.039}_{-0.029}^{+0.019}$ & $p_0 > p_1$ & 93\% \\
Kepler-39 & ${0.0984}_{-0.0074}^{+0.0059}$ & ${0.08813}_{-0.00041}^{+0.00062}$ & ${0.20}_{-0.15}^{+0.08}$ & $p_0 > p_1$ & 90\% \\
Kepler-412 & ${0.0909}_{-0.0065}^{+0.0083}$ & ${0.1028}_{-0.0013}^{+0.0015}$ & ${-0.28}_{-0.22}^{+0.22}$ & $p_0 < p_1$ & 90\% \\
GJ 1214 & ${0.136}_{-0.059}^{+0.028}$ & ${0.1101}_{-0.0015}^{+0.0021}$ & ${0.3}_{-1.4}^{+0.1}$ & $p_0 > p_1$ & 71\% \\
HD 80606 & ${0.053}_{-0.019}^{+0.029}$ & ${0.1017}_{-0.0010}^{+0.0010}$ & ${-2.6}_{-5.8}^{+2.1}$ & $p_0 < p_1$ & 96\% \\
TRAPPIST-1 b & ${0.074}_{-0.015}^{+0.017}$ & ${0.0859}_{-0.0007}^{+0.0006}$ & ${-0.18}_{-0.70}^{+0.29}$ & $p_0 < p_1$ & 74\% \\
TRAPPIST-1 c & ${0.0824}_{-0.0095}^{+0.0089}$ & ${0.08709}_{-0.00058}^{+0.00057}$ & ${-0.11}_{-0.31}^{+0.20}$ & $p_0 < p_1$ & 70\% \\
\end{tabular}
\caption{Maximum-likelihood parameters for $p_0, p_1$, the self-contamination parameter $\epsilon = 1 - (p_1/p_0)^2$, and the confidence interval (CI) for each detection of $p_0$ relative to $p_1$.} \label{tab:results}
\end{table*}


\section{Discussion} \label{sec:discussion}

We have investigated the prospects for robust measurement the radius ratios of planets using
the ratio of the ingress duration to total transit duration, which should be
less affected by contamination by starspots than the transit depth measurements.
We have then applied this to a sample of ten stars to look for differences between
the radius ratio measured from ingress ($p_0$) versus that measured from transit
depth ($p_1$).

\begin{figure}
    \centering
    \includegraphics[scale=0.45]{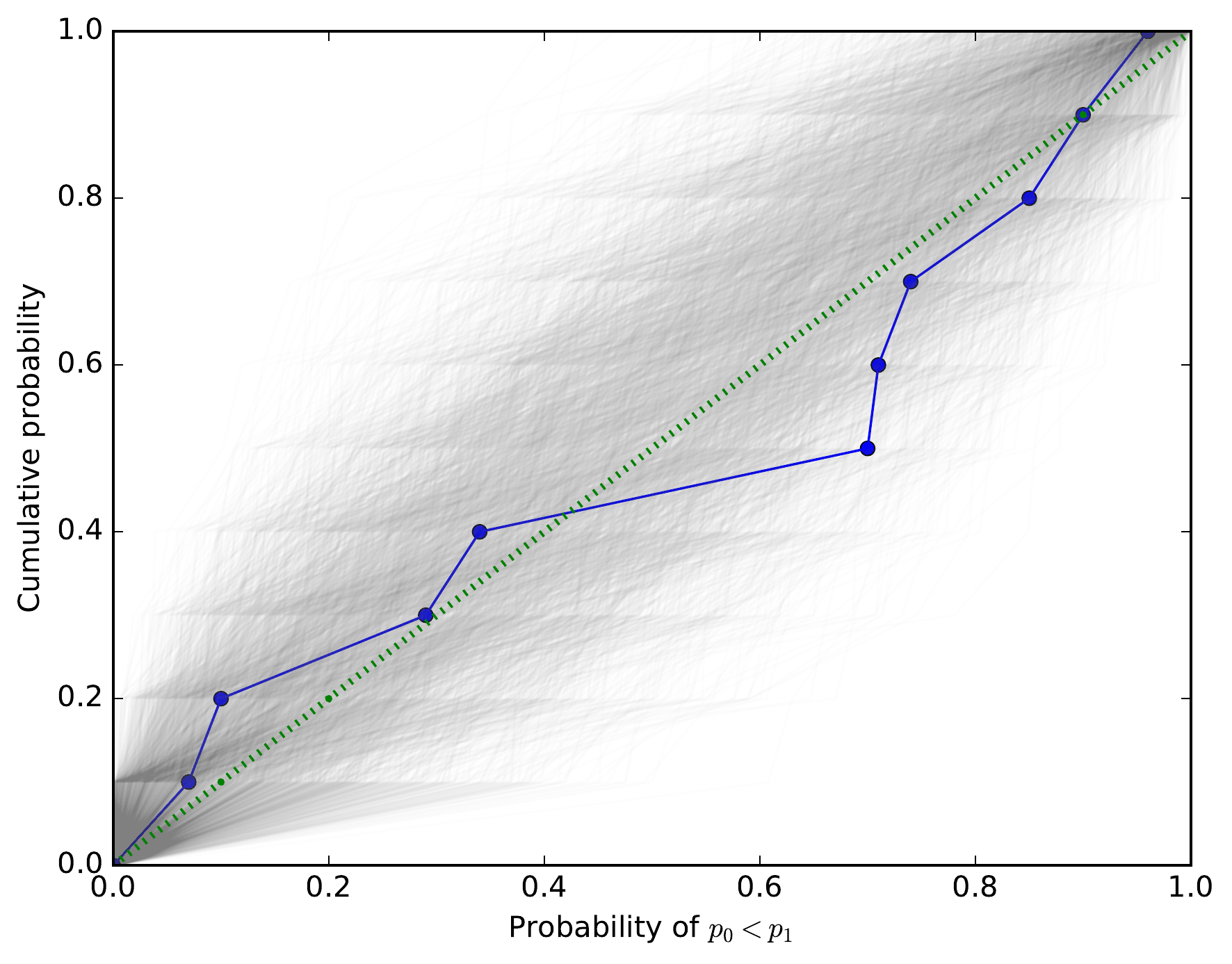}
    \caption{Cumulative probability distribution for the probability that
    $p_0 < p_1$ for the ten stars in our sample (blue points).  The light grey lines show $3000$ Monte
    Carlo simulations of probabilities drawn from a uniform distribution between zero
    and one.  The green curve shows a uniform probability distribution.}
    \label{fig:ks_sample}
\end{figure}

Given the probability measurements of $p_0 < p_1$ for the ten stars in our study, we can ask the question of
whether there is evidence that the probability sample is drawn from a uniform probability distribution.  
We sort the probabilities of $p_0 < p_1$ for the ten stars in our study, and plot these versus a uniform
probability distribution.  Figure \ref{fig:ks_sample} shows the cumulative probability distribution 
of the sample versus the measured probabilities.  If the distributions of $p_0$ and $p_1$ are
statistically consistent for the entire sample, then we expect that ten measured probabilities
of $p_0 < p_1$ will be consistent with being drawn from a uniform probability distribution.  
We apply the
K-S test to the sample, finding a maximum distance of $-0.2$ for a sample size of 10,
giving a probability that this is drawn from a uniform distribution of 28\%.  We view this
as indicating that the distribution of $p_0<p_1$ probabilities is likely drawn from a uniform 
distribution;  i.e. that the sample as a whole is consistent with $p_0 = p_1$ for all planets.

Similar analyses have been conducted for the single systems CoRoT-2 \citep{Bruno2016} and CoRoT-7 \citep{Barros2014}, and more general discussions appear in \citet{Oshagh2013} and \citet{Csizmadia2013}. The main differences between this work and others is that we explicitly use the ingress and egress durations to measure the true planet radii in the presence of bright or dark regions on the stellar surface.

This work echoes \citet{Csizmadia2013} -- in particular, we found that fixing the limb-darkening parameters to their theoretical values for several systems introduced artificial constraints on the impact parameter and $p_1$, occasionally producing apparent discrepancies where $p_0 \neq p_1$, which become consistent once the limb-darkening parameters are allowed to float. We encourage users of this light curve parameterization to fit for the limb-darkening parameters whenever possible, though we note the exceptionally good fit to TrES-2 using the quadratic limb-darkening parameters of \citep{Magic2015}.

We also note the importance of asteroseismic constraints on the orbital parameters for several systems. The asteroseismic stellar density provides us with an independent measurement of $a/R_\star$, which is otherwise degenerate with $i_o, p_0$ and the limb-darkening parameters. Bright TESS targets may also be amenable to asteroseismic density measurements, and therefore good candidates for this self-contamination analysis. Alternatively the stellar density can be measured to high precision in multi-planet systems from the planet periods and transit durations, which will be valuable for dim stars like TRAPPIST-1.

Diminished limb-darkening in the infrared promises that \spitzer and JWST observations of transiting exoplanets will be interesting subjects for analysis with the self-contamination analysis. Ground-based follow-up may also be a fruitful means of detecting the affects of stellar activity on transiting exoplanet light curves, especially with the high precision and time resolution enabled by holographic diffusers \citep[see e.g.][]{Stefansson2017, Morris2018d}.

\section{Conclusions} \label{sec:conclusion}

We have presented a reparameterization of the transit light curve of \citet{Mandel2002} which splits the original ``$p$'' parameter into two parameters: $p_0 = R_p/R_\star$ which defines the planet radius, and $p_1 \approx \sqrt{\delta}$ which defines the transit depth. This parameterization allows the transit model to account for significant contamination by bright or dark active regions on the stellar surface, or significant planetary oblateness in the planet's direction of motion. The resulting constraint on $p_0 = R_p/R_\star$ is more accurate in the presence of stellar activity (and usually less precise).

We fit the light curves of several transiting planets to study the photospheric inhomogeneities of their host stars, see Table~\ref{tab:results}. We report $p_0 \approx p_1$ (no significant detection of self-contamination) for \kepler light curves of TrES-2, HAT-P-7, TRAPPIST-1 and \spitzer observations of GJ 1214.  We find that the uncertainties on $p_0$ are typically an order of magnitude larger than $p_1$, consistent with estimates of the noise (Appendix C).
We find little evidence for transit depth dilution due to occulted starspots ($p_0 > p_1$) for the well-studied spotted pair of host stars HAT-P-11 and Kepler-17, and we find weak evidence that the transiting planets of Kepler-412 and HD 80606 likely occult relatively bright regions of the stellar photosphere ($p_0 < p_1$). We can recover the reported oblateness of Kepler-39 b using this parameterization, though its detection is not statistically significant.   We note that in general the oblateness is likely degenerate with self-contamination -- so the unusually large inferred oblateness may have an alternate explanation as being due to an inhomogeneous photosphere.  

In the best cases we have studied in this paper, HAT-P-11b, Kepler-17b, and TrES-2b, we obtain sensitivity to
self-contamination, $\epsilon$ at the 1-3\% level (Table \ref{tab:results}), which is close enough to the
expected levels of contamination that we were unable to achieve a definitive detection.
This ratio of precisions on $p_0$ range from $5-25$ times the uncertainties on $p_1$ (Table \ref{tab:results}) 
The main advantage, then, in measuring $p_0$ is to obtain the radius ratio of the planet
to the star when the unknown contamination of the stellar flux is significant.
The other circumstance which may favor measuring $p_0$ is when there is significant
red noise which has a larger amplitude on the timescale of the transit than on the
timescale of ingress/egress.  
Granulation noise tends to be strongly red, and thus will affect the measurement
of the transit depth more strongly than the measurement of the ingress/egress duration.

With the James Webb Space Telescope (JWST) we expect to obtain much higher precision on $p_0$ due to the larger collecting area.
As an example, the Trappist-1 system has ingress/egress durations of $\approx 2-5$ minutes, and when
observed with NIRSPEC, we expect precisions of $2-5$ seconds for the precision of these measurements.
This will enable a measurement of $p_0$ comparable to the current precisions obtained with Spitzer
for which multiple transits have been observed for each planet.  Thus, we expect that JWST will yield
constraints upon $p_0$ which will give accurate radius ratios for these planets to diagnose the contamination by starspots.  We find our simulations with extremely high signal-to-noise (\S \ref{sec:valid}) demonstrate that the radius-ratio of the planet measured from $p_0$ can be recovered more accurately than the radius ratio measured from $p_1$ in the presence of star spot contamination, since $p_1$ is biased by the presence of star spots (although $p_1$ is measured more precisely).

We have shown that $p_0$ is degenerate with the limb-darkening parameters. As a result, this technique will work better at red wavelengths, where limb-darkening is less severe, than in the blue. However, the contrast of starspots against the stellar photosphere diminishes as one observes at longer wavelengths as well, so this technique might be best-suited to observations in the red optical, where the spot contrast may be significant but the limb-darkening is weaker.

This technique is complementary to other starspot measurement techniques -- it is sensitive to time-independent spots, though it  provides a weaker signal than techniques like the flux deficit. The flux deficit technique -- which measures variations in flux as a star rotates -- has been applied to vast numbers of \kepler stars \citep{Walkowicz2013, Notsu2013, Mathur2014}. Or alternatively, spot distributions can be revealed by detailed modeling of individual spots throughout rotational modulation \citep{Davenport2015}, or via observations of spot occultations by transiting exoplanets \citep[for example: ][]{Wolter2009,Sanchis-Ojeda2011,Davenport2015thesis,Morris2017a,Dai2018}.

\acknowledgements

We thank Laura Kreidberg, Ian Dobbs-Dixon, Travis Berger, and Dan Huber.  We acknowledge support from NSF grant 1615315. 

This research has made use of NASA's Astrophysics Data System. This research has made use of the NASA Exoplanet Archive, which is operated by the California Institute of Technology, under contract with the National Aeronautics and Space Administration under the Exoplanet Exploration Program. Some of the data presented in this paper were obtained from the Mikulski Archive for Space Telescopes (MAST). STScI is operated by the Association of Universities for Research in Astronomy, Inc., under NASA contract NAS5-26555. This paper includes data collected by the Kepler mission. Funding for the Kepler mission is provided by the NASA Science Mission directorate. This paper includes data collected by the K2 mission. Funding for the K2 mission is provided by the NASA Science Mission directorate. This work is based in part on observations made with the Spitzer Space Telescope, which is operated by the Jet Propulsion Laboratory, California Institute of Technology under a contract with NASA. This research has made use of the VizieR catalogue access tool, CDS, Strasbourg, France. The original description of the VizieR service was published in A\&AS 143, 23. This work has made use of data from the European Space Agency (ESA) mission {\it Gaia} (\url{https://www.cosmos.esa.int/gaia}), processed by the {\it Gaia} Data Processing and Analysis Consortium (DPAC, \url{https://www.cosmos.esa.int/web/gaia/dpac/consortium}). Funding for the DPAC has been provided by national institutions, in particular the institutions participating in the {\it Gaia} Multilateral Agreement.

\software{\texttt{batman} \citep{Kreidberg2015}, \texttt{emcee} \citep{Foreman-Mackey2013}, \texttt{corner} \citep{Foreman-Mackey2016}, \texttt{astropy} \citep{Astropy2018}, \texttt{ipython} \citep{ipython}, \texttt{numpy} \citep{VanDerWalt2011}, \texttt{scipy} \citep{scipy},  \texttt{matplotlib} \citep{matplotlib}}

\facilities{\kepler, K2, \spitzer, ESA/Gaia}

\bibliography{bibliography}
\bibliographystyle{aasjournal}

\appendix

\section{Planet radius from the duration of ingress/egress} \label{sec:simpleeqns}

For a transiting exoplanet in a circular orbit, the transit duration from first through fourth contact is
\begin{equation}
T _ {14} = \frac { P } { \pi } \sin ^ { - 1} \left[ \frac { R _ { \star } } { a } \frac { \sqrt { ( 1+ p_0 ) ^ { 2} - b ^ { 2} } } { \sin i } \right],
\end{equation}
and the duration from second through third contact is
\begin{equation}
T _ {23} = \frac { P } { \pi } \sin ^ { - 1} \left[ \frac { R _ { \star } } { a } \frac { \sqrt { ( 1- p_0 ) ^ { 2} - b ^ { 2} } } { \sin i } \right],
\end{equation}
where $p_0 = R_p/R_\star$, $P$ is the orbital period, $b$ is the impact parameter, $i$ is the planet's orbital inclination, and $a/R_\star$ is the scaled orbital semi-major axis \citep{Winn2011}. We can invert  these relations to find the planet-to-stellar radius ratio as a function of the transit durations,
\begin{equation}
\frac{R_p}{R_\star} = \frac{1}{2} \left( \left[ \left[ \frac{a}{R_\star}\sin{\frac{\pi T_{14}}{P}}\sin{i}\right]^2 + b^2\right]^{1/2} - \left[ \left [ \frac{a}{R_\star}\sin{\frac{\pi T_{23}}{P}}\sin{i}\right]^2 + b^2 \right ]^{1/2} \right).
\end{equation}
We can get a sense for the scaling of the terms in this equation by simplifying to the case where $i=90^\circ$, $b=0$ and $T_{14}, T_{23} \ll P$ using the small angle approximation,
\begin{equation}
\frac{R_p}{R_\star} \sim \frac{\pi}{2P} \frac{a}{R_\star} \left( T_{14} - T_{23}\right).
\end{equation}
The constraint on the planetary radius ($p_0=R_p/R_\star$ in our reparameterization) primarily comes from the difference between the durations of first-through-fourth contact and second-through-third contact. 

\section{Limiting cases: Extreme spot distributions} \label{sec:limitingcases}

In Sections~\ref{sec:activelat} and \ref{sec:brightlat} we assumed that the starspots had the same contrast as sunspots, $c=0.7$. In this section, we revisit those toy-model analyses with more extreme spot contrasts, by setting the contrast to that of sunspot umbrae, $c=0.2$. 

\begin{figure*}
    \centering
    \includegraphics[scale=0.5]{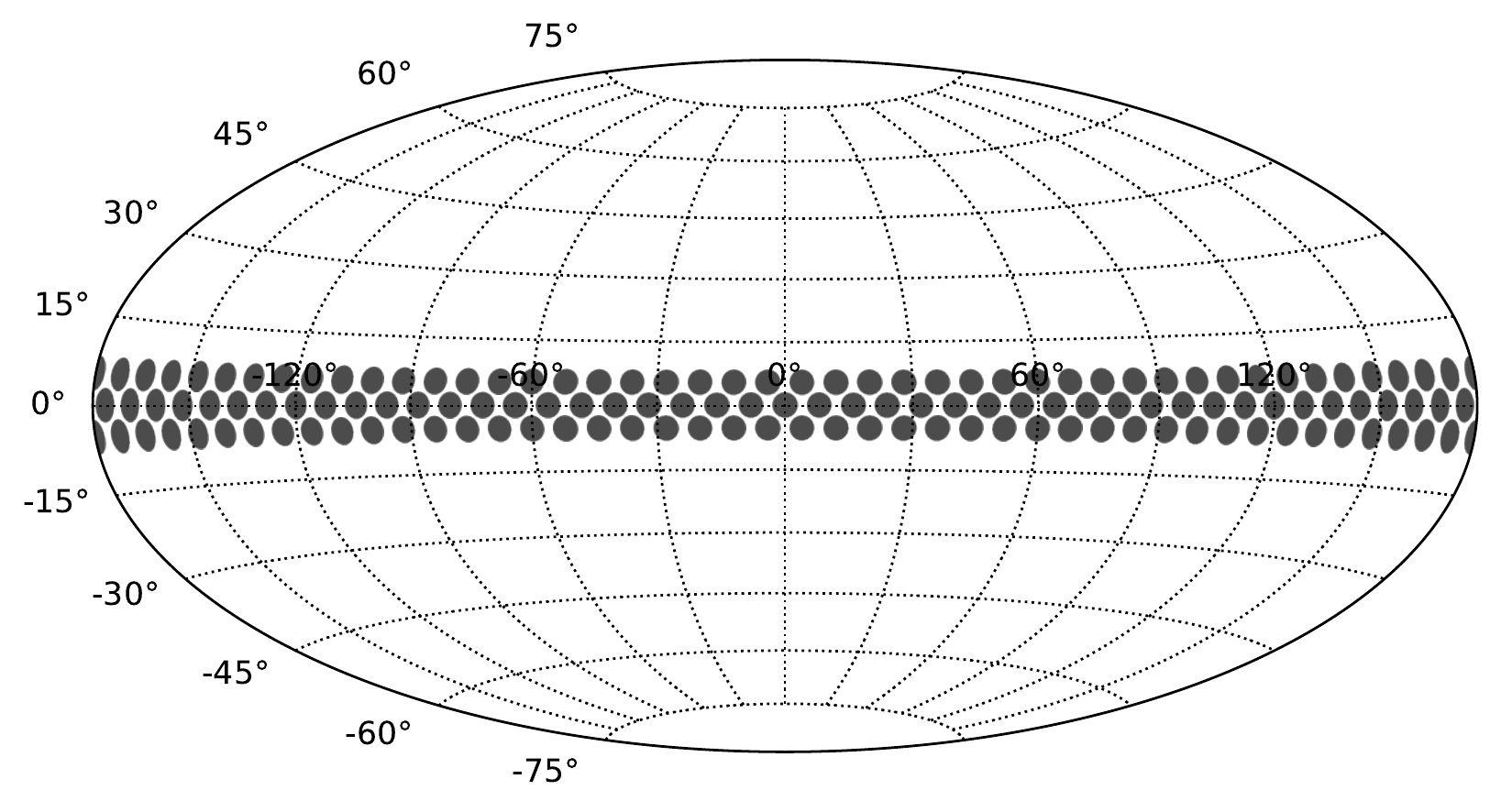}
    \includegraphics[scale=0.6]{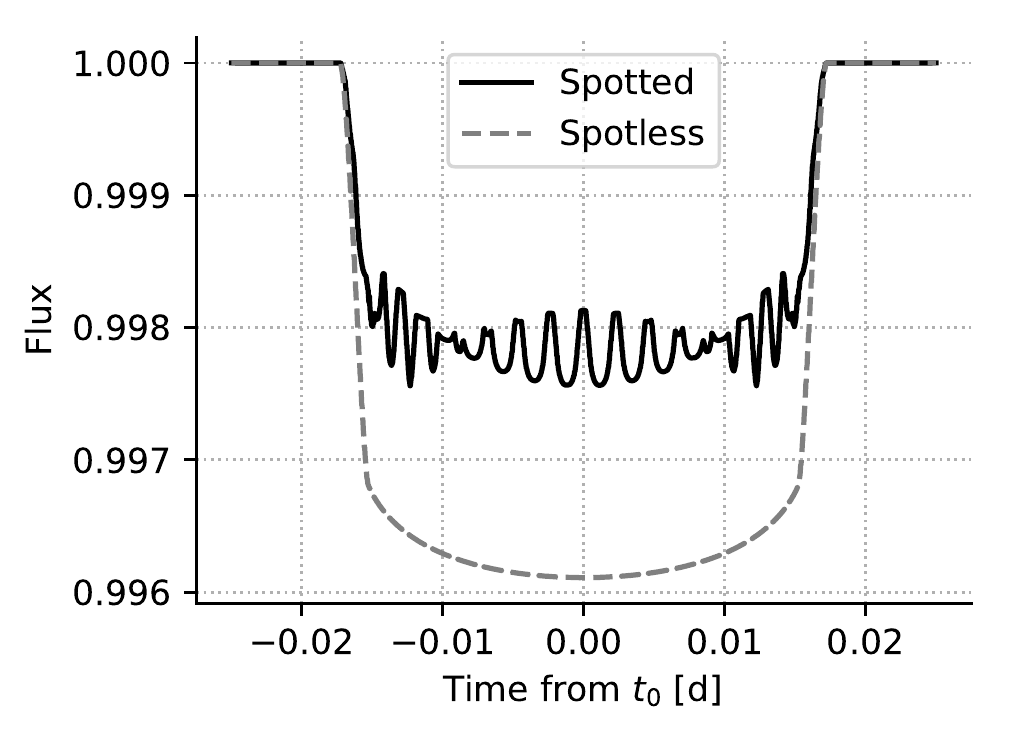}
    \includegraphics[scale=0.7]{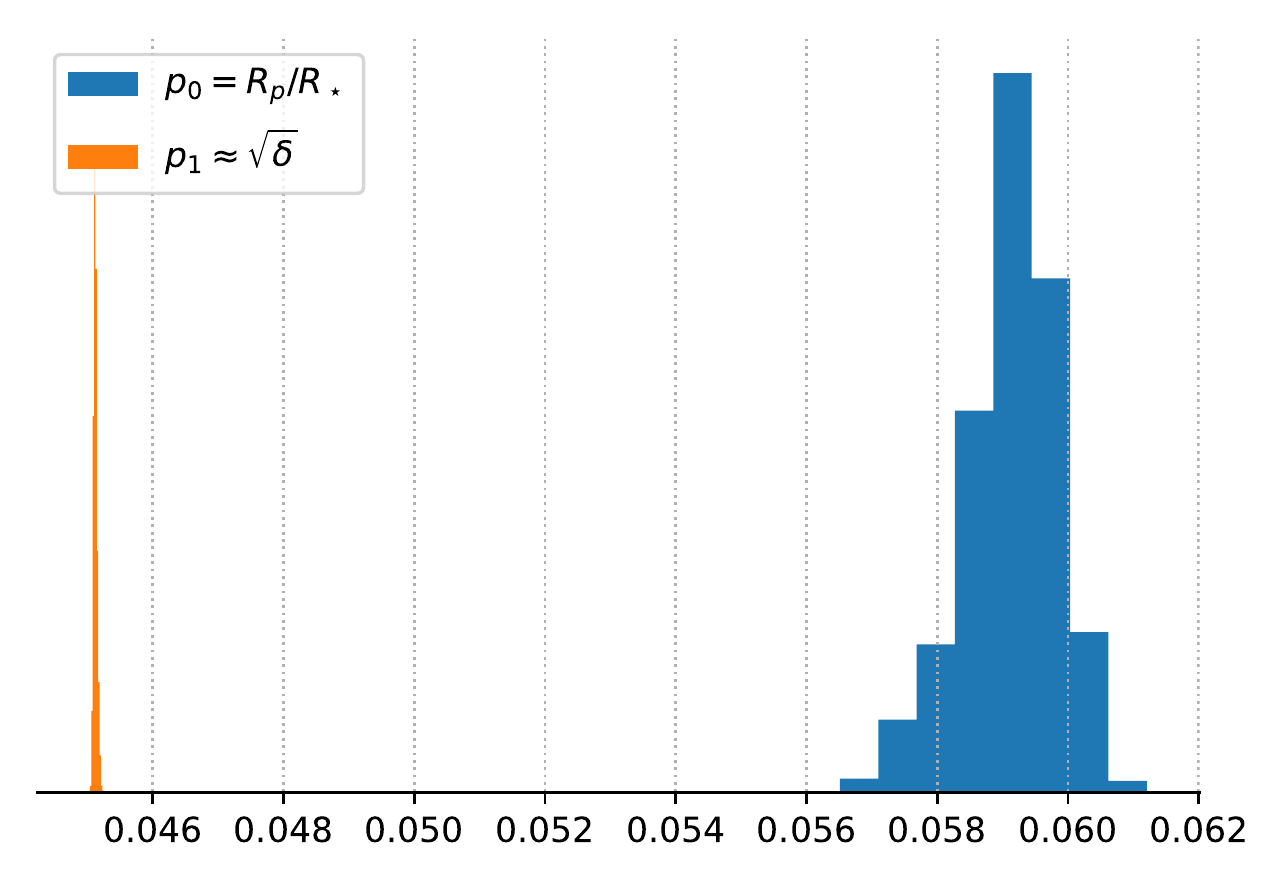}
    \caption{\textsl{Left:} Hypothetical spot map for a dense band of spots in an active latitude at the stellar equator, with contrast $c=0.2$. \textsl{Right:} Transit of a small planet across the star with the spot map on the left, with $i_o = 90^\circ$ and $\lambda = 0^\circ$ and other parameters set to those of TRAPPIST-1 g (black curve), compared with the transit of the same system without spots (gray dashed curve). At lower S/N, or for different spot geometries, the bottom of the spotted transit might simply appear flat and relatively shallow, but the ingress and egress durations are the same for both light curves, allowing us to recover the true planet radius from timing, independent of the transit depth which is affected by starspots.}
    \label{fig:map4}
\end{figure*}

First we simulate a dense band of spots with contrast $c=0.2$ centered on the stellar equator -- see Figure~\ref{fig:map4}. The self-contamination parameter $\epsilon = 1 - (p_1/p_0)^2 = -0.598_{-0.032}^{+0.014}$ in this limiting case. 

\begin{figure*}
    \centering
    \includegraphics[scale=0.5]{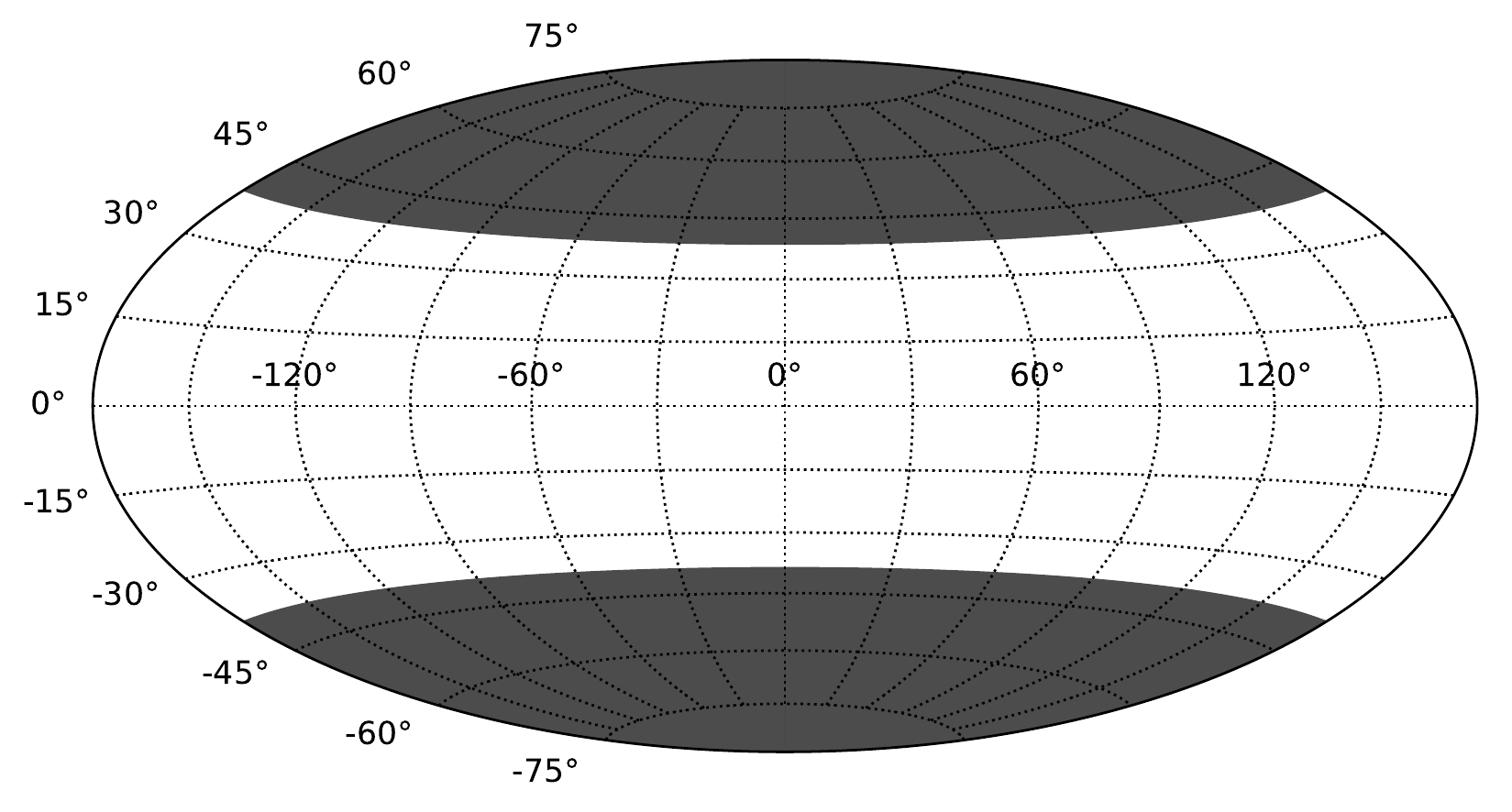}
    \includegraphics[scale=0.6]{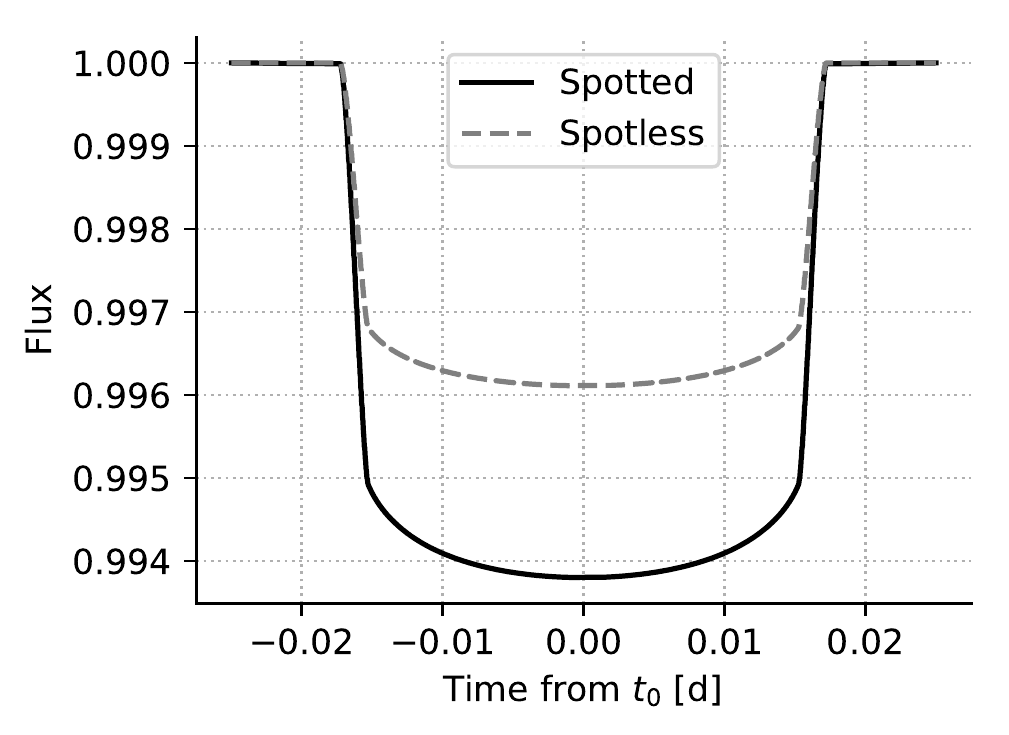}
    \includegraphics[scale=0.7]{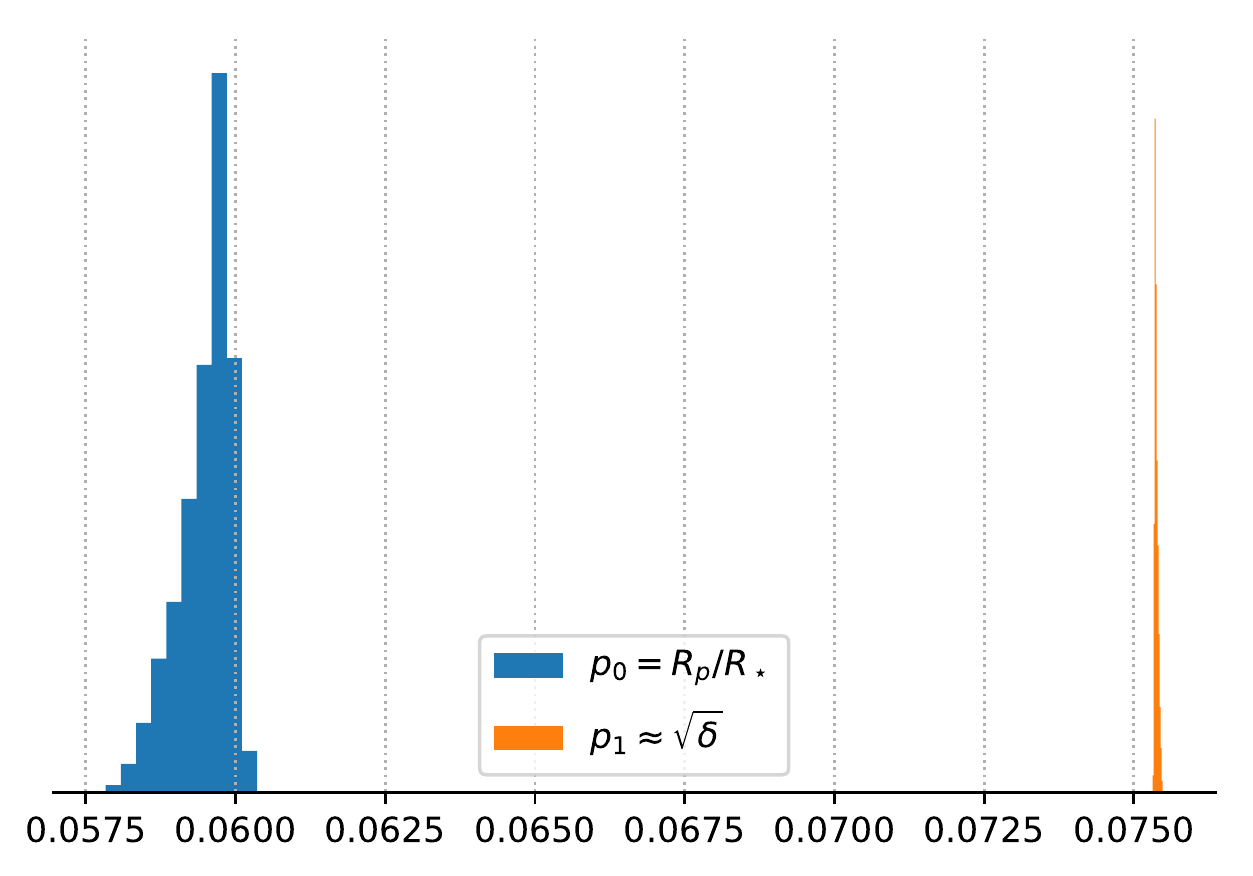}
    \caption{\textsl{Left:} Hypothetical spot map for dark polar spots that cover much of the star except the transit chord, with contrast $c=0.2$. \textsl{Right:} Transit of a small planet across the star with the spot map on the left, with $i_o = 90^\circ$ and $\lambda = 0^\circ$ and other parameters set to those of TRAPPIST-1 g (black curve), compared with the transit of the same system without spots (gray dashed curve). At lower S/N, or for different spot geometries, the bottom of the spotted transit might simply appear flat and relatively shallow, but the ingress and egress durations are the same for both light curves, allowing us to recover the true planet radius from timing, independent of the transit depth which is affected by starspots.}
    \label{fig:map3}
\end{figure*}

Next we simulate a pair of polar spots that extend almost all the way from the pole to the edge of the transit chord, with contrast $c=0.2$ -- see Figure~\ref{fig:map3}. The self-contamination parameter $\epsilon = 0.419_{-0.015}^{+0.011}$ in this case. 

Thus one might expect $\epsilon$ to vary roughly on $-0.5 \lesssim \epsilon \lesssim 0.5$ for extreme self-contamination due to occulted or unocculted starspots. 

\section{Uncertainties}

Using the estimates by \citet{Carter2008} for the uncertainties on the ingress and total duration given Fisher information analysis of a piecewise-linear transit model, we can estimate the uncertainties on $p_0$ and $p_1$. For flux observations with independent Gaussian uncertainties $\sigma$, for a transit with depth $\delta$, ingress duration $\tau$, mid-ingress to mid-egress duration $T$, defined as 
\begin{eqnarray}
T &\equiv& 2 \tau_0 \sqrt{1-b^2} \\ 
\tau &\equiv& 2 \tau_0 \frac{p_0}{\sqrt{1-b^2}} \\ 
\end{eqnarray}
where $\tau_0$ is
\begin{equation}
\tau_0 = \frac{P}{2\pi}\frac{R_\star}{a},
\end{equation}
with uncertainties on $\tau$ and $T$ given by
\begin{eqnarray}
\sigma_\tau &\approx& \frac{\sigma}{\delta T} \sqrt{\frac{6 \tau}{\Gamma}} ,\\
\sigma_T &\approx& \frac{\sigma}{\delta T} \sqrt{\frac{2 \tau}{\Gamma}},
\end{eqnarray}
where $\Gamma$ is the sampling rate (if $N$ photometric measurements are
taken uniformly over a total observing time $T_0$, then $\Gamma = N/T_0$),
so $\Gamma^{-1}$ is the time between the start of successive exposures.
Rearranging, we find
\begin{equation}
p_0 = \frac{\tau T}{4\tau_0^2}, 
\end{equation}
and therefore the uncertainty, $\sigma_{p_0}$, is given by 
\begin{equation}
\left(\frac{\sigma_{p_0}}{p_0}\right)^2 = \frac{T}{Q^2} \left[\frac{6+\eta}{\tau} \right]+\left(\frac{\sigma_{\tau_0}}{\tau_0}\right)^2
\end{equation}
in the limit of $\tau \ll T$, where $\eta = T/(T_0-T-\tau)$ is the ratio of the duration of photometric measurements in-transit to out-of-transit ($\eta \rightarrow 0$ for perfectly known out-of-transit flux). From this equation we see conclusions we discussed elsewhere in this work, for example, if $\tau_0$ is known imprecisely by lack of prior constraints on the stellar density or transit impact parameter, then $p_0$ will have large uncertainties. 

If we know the stellar density/impact parameter well ($\sigma_{\tau_0}\rightarrow 0$) and measure the out-of-transit flux well ($\eta \rightarrow 0$), then 
\begin{equation} \label{error_p0}
\frac{\sigma_{p_0}}{p_0} = \frac{\sigma 6^{1/2}}{p_1^2(\Gamma \tau)^{1/2}} \propto \frac{(1-b^2)^{1/4}}{p_0^{1/2}},
\end{equation}
i.e., we obtain higher precision on $p_0$ for systems with higher impact parameters, for which the ingress/egress durations are longer. 

The uncertainty on $p_1$ can be computed from $\sigma_\delta$ given $p_1^2 \approx \delta$, so 
\begin{equation}
\sigma_{p_1} = \frac{\sigma_{p_1^2}}{2 p_1}
\end{equation}
thus 
\begin{equation} \label{error_p1}
\frac{\sigma_{p_1}}{p_1} = \frac{\sigma}{2 p_1^2 (\Gamma T)^{1/2}}.
\end{equation}

Comparing the fractional uncertainties on $p_0$ (eqn \ref{error_p0}) and $p_1$ (eqn \ref{error_p1}),
we find:
\begin{equation}
\frac{\sigma_{p_0}}{p_0} \left(\frac{\sigma_{p_1}}{p_1}\right)^{-1} = 24^{1/2} \left(\frac{T}{\tau}\right)^{1/2} = 15.5 (1-b^2)^{1/2} (p_0/0.1)^{-1/2}.  
\end{equation}
Consequently, the fractional uncertainty on $p_0$ tends to be $\approx$ an order
of magnitude larger than $p_1$; this can be exacerbated further by strong limb-darkening
which causes the depth of transit at ingress and egress to be shallower than the
mean transit depth. 

The foregoing equations assume uncorrelated, time-independent 
white noise, which will not be the case for high precision measurements at which
point the uncertainties will be dominated by stellar granulation variability.

\end{document}